\documentclass[aps,prb,twocolumn,superscriptaddress,showpacs]{revtex4}
\usepackage{graphicx,epsfig}
\usepackage{dcolumn}
\usepackage{bm}
 \begin{document}
\title{Plasmon mass and Drude weight in strongly spin-orbit-coupled 2D electron gases}
\author{Amit Agarwal}
\email{amit.agarwal@sns.it}
\affiliation{NEST, Istituto Nanoscienze-CNR and Scuola Normale Superiore, I-56126 Pisa, Italy}
\author{Stefano Chesi}
\affiliation{Department of Physics, University of Basel, Klingelbergstrasse 82, 4056 Basel, Switzerland}
\author{T. Jungwirth}
\affiliation{Institute of Physics ASCR, v.v.i., Cukrovarnick 10, 162 53 Praha 6, Czech Republic}
\affiliation{School of Physics and Astronomy, University of Nottingham, Nottingham NG7 2RD, United Kingdom}
\author{Jairo Sinova}
\affiliation{Department of Physics, Texas A\&M University, College Station, Texas 77843-4242, USA}
\affiliation{Institute of Physics ASCR, v.v.i., Cukrovarnick 10, 162 53 Praha 6, Czech Republic}
\affiliation{Kavli Institute for Theoretical Physics China, CAS, Beijing 100190, China}
\author{G. Vignale}
\affiliation{Department of Physics and Astronomy, University of Missouri, Columbia, Missouri 65211, USA}
\affiliation{Kavli Institute for Theoretical Physics China, CAS, Beijing 100190, China}
\author{Marco Polini}
\email{m.polini@sns.it} \homepage{http://qti.sns.it}
\affiliation{NEST, Istituto Nanoscienze-CNR and Scuola Normale Superiore, I-56126 Pisa, Italy}
\affiliation{Kavli Institute for Theoretical Physics China, CAS, Beijing 100190, China}

\begin{abstract}
Spin-orbit-coupled two-dimensional electron gases (2DEGs) are a textbook example of {\it helical~Fermi~liquids}, {\it i.e.} quantum liquids in which spin (or pseudospin) and momentum degrees-of-freedom at the Fermi surface have a well-defined correlation. Here we study the long-wavelength plasmon dispersion and the Drude weight of archetypical spin-orbit-coupled 2DEGs. We first show that these measurable quantities are sensitive to electron-electron interactions due to broken Galileian invariance and then discuss in detail why the popular random phase approximation is not capable of describing the collective dynamics of these systems even at very long wavelengths. This work is focussed on presenting approximate microscopic calculations of these quantities based on the minimal theoretical scheme that captures the basic physics correctly, {\it i.e.} the  time-dependent Hartree-Fock approximation. We find that interactions enhance the ``plasmon mass" and suppress the Drude weight. Our findings can be tested by inelastic light scattering, electron energy loss, and far-infrared optical-absorption measurements.
\end{abstract}
 \maketitle
\section{Introduction}
\label{sect:intro}

In recent years we have witnessed a tremendous explosion of interest in a large variety of novel two-dimensional (2D) quantum many-body systems. 
Prime examples are: (i) strongly spin-orbit-coupled 2D electron and hole gases, which are promising candidates for semiconductor 
spintronics~\cite{soc_reviews}; (ii) graphene~\cite{graphene_review} (a monolayer of carbon atoms arranged in a 2D honeycomb lattice), which has attracted a great deal of interest because of the massless-Dirac-fermion character of its carriers and because it may pave the way for carbon-based electronics~\cite{avouris_natnano_2007}; (iii) 2D electron gases in HgTe/Hg(Cd)Te quantum wells where massless Dirac fermions are predicted to arise at a critical quantum well thickness~\cite{bernevig_science_2006,konig_science_2007,konig_jpsp_2008,brune_natphys_2010}; and, more recently, (iv) metallic surface states of 3D topological insulators~\cite{TI_reviews,xia_natphys_2009,zhang_natphys_2009,zhang_prl_2009}. 

These systems share a unique common factor: their orbital degrees-of-freedom are intimately coupled to the electron spin (or sublattice pseudospin, in the case of graphene) degree-of-freedom. This coupling, being of relativistic origin~\cite{footnote_crystal_field_effect}, naturally breaks Galileian invariance and is thus the basic reason for a quite sensitive dependence of several observables to electron-electron interactions, even at very long wavelengths (see for example Refs.~\onlinecite{shekhter_prb_2005,farid_prl_2006,polini_arXiv_2009,principi_prb_2009,tse_prb_2009,abedinpour_arxiv_2010}). Furthermore, these systems exhibit coupled spin-charge collective dynamics, which is just beginning to be investigated in the contemporary literature~\cite{raghu_prl_2010}. 

In this article we focus our attention on an archetypical 2D electron gas model Hamiltonian with spin-orbit coupling (SOC). For the sake of simplicity we choose an elementary form of SOC which is linear in momentum and has the canonical Rashba {\it or} Dresselhaus functional form. Since collective dynamics in quantum many-body systems is controlled by isolated poles in dynamical linear-response functions~\cite{Giuliani_and_Vignale}, we carry out a microscopic study of the density-density response function in the dynamical limit taking into account exactly SOC and treating electron-electron interactions beyond the random phase approximation (RPA). The RPA, which is commonly used to describe electron liquids, is indeed not capable to capture the subtle renormalization of the plasmon mode that occurs in non-Galileian-invariant quantum liquids. The study of many-body effects when {\it both} Rashba and Dresselhaus SOC terms are present in the Hamiltonian is beyond the scope of the present article: rotational invariance of the Fermi contours is indeed spoiled by the simultaneous presence of both effects and this complicates (and partly obscures) the basic interplay between SOC and many-body effects we want to highlight. 
Electron-electron interactions in 2D electron and hole gases in the presence of SOC have attracted a certain deal of attention~\cite{chen_prb_1999,chen_secondprb_1999,magarill_jept_2001,mishchenko_prb_2003,shekhter_prb_2005,saraga_prb_2005,wang_prb_2005,dimitrova_prb_2005,pletyukhov_prb_2006,schliemann_prb_2006,alvaro_unpublished_2006,farid_prl_2006,pletyukhov_epjb_2007,chesi_prb_2007,chao_apl_2008,ambrosetti_prb_2009,badalyan_prb_2009,badalyan_prb_2010,nechaev_prb_2010,zak_prb_2010,chesi_arxiv_2010,chesi_arxiv_2010_bis}. Below we will make contact with the pre-existing literature whenever possible. 

Our manuscript is organized as follows. In Sect.~\ref{sect:theory} we present the model we have studied, we introduce the basic definitions, and outline the equation-of-motion approach we have used to relate the density-density response function with the longitudinal current-current response function. The latter is then evaluated microscopically within the time-dependent Hartree-Fock approximation in the long-wavelength limit in Sect.~\ref{sect:TDHFtheory}. While the main focus of this paper is on the plasmon dispersion at long wavelengths and on the Drude weight, in the same Section we briefly discuss interaction corrections to the spin Hall conductivity and the renormalization of the spin-orbit splitting of the bands due to electron-electron interactions. In Sect.~\ref{sect:numerics} we present our main numerical results, while in Sect.~\ref{sect:conclusions} we summarize our findings and draw our main conclusions.

\section{General theory}\label{sect:theory}
\subsection{Model Hamiltonian}
\label{sect:hamil}

We consider the following model Hamiltonian for a 2D electron gas (2DEG),
\begin{equation}\label{eq:Hamiltonian}
{\hat {\cal H}} = {\hat {\cal H}}_0 + {\hat {\cal H}}_{\rm SOC} + {\hat {\cal H}}_{\rm int}~,
\end{equation}
incorporating the usual parabolic-band kinetic-energy term, SOC, and electron-electron interactions. More precisely, the first term in Eq.~(\ref{eq:Hamiltonian}) is given by
\begin{equation}\label{eq:kin}
{\hat {\cal H}}_0 = \sum_{{\bm k}, i} \varepsilon(k){\hat \psi}^\dagger_{{\bm k}, i} {\hat \psi}_{{\bm k}, i}~,
\end{equation}
with $i = \uparrow, \downarrow$  a real-spin label and $\varepsilon(k) = \hbar^2 k^2/(2m_{\rm b})$, $m_{\rm b}$ being the bare electron band mass. For the SOC term we choose a simple linear-in-momentum Rashba-Dresselhaus model~\cite{winkler_book}:
\begin{eqnarray}\label{eq:SOC}
{\hat {\cal H}}_{\rm SOC} &=& \sum_{{\bm k}, i, j} 
{\hat \psi}^\dagger_{{\bm k}, i} [\alpha (\sigma^x_{ij}k_y -\sigma^y_{ij} k_x) \nonumber \\
&+& \beta (\sigma^x_{ij}k_x -\sigma^y_{ij} k_y)] {\hat \psi}_{{\bm k}, j}~.
\end{eqnarray}
Here $\sigma^x_{ij}$ and $\sigma^y_{ij}$ are Pauli matrices, while $\alpha$ and $\beta$ are the Rashba and Dresselhaus SOC constants, respectively. Diagonalization of the sum of the first two terms in Eq.~(\ref{eq:Hamiltonian}) 
(${\hat {\cal H}}_0+{\hat {\cal H}}_{\rm SOC}$) yields two bands (see, for example, Ref.~\onlinecite{trushin_prb_2009}),
\begin{equation}\label{eq:bands}
\varepsilon_\lambda({\bm k}) = \varepsilon(k) +\lambda k \Gamma({\theta_{\bm k}})~,
\end{equation}
with $\lambda = \pm 1$ the so-called ``chirality" index, $\theta_{\bm k}$ the angle between ${\bm k}$ and the ${\hat {\bm x}}$ axis, and
\begin{equation}\label{eq:gamma}
\Gamma(\theta) = \sqrt{\alpha^2+\beta^2+4\alpha\beta\sin(\theta)\cos(\theta)}~.
\end{equation}
The Fermi wave vectors for the two bands can be expressed in terms of $\theta_{\bm k}$ and of the Fermi energy $\varepsilon_{\rm F}$:
\begin{equation}\label{eq:kF}
k_{{\rm F}, \lambda}^{(0)} = -\lambda~\frac{m_{\rm b} \Gamma(\theta_{\bm k}) }{\hbar^2}
+ \sqrt{\left[\frac{m_{\rm b} \Gamma(\theta_{\bm k})}{\hbar^2}\right]^2 + 
\frac{2m_{\rm b} \varepsilon_{\rm F}}{\hbar^2}}~.
\end{equation}
Note that for zero Fermi energy the Fermi contour of the minority $\lambda = +$ band
shrinks into a single point ({\it i.e.} $k_{{\rm F}, +}^{(0)}=0$). For any $\varepsilon_{\rm F}\ge 0$ 
the electron density $n$ can be expressed in terms of the Fermi energy as
\begin{equation}\label{eq:density}
n = \frac{m_{\rm b} \varepsilon_{\rm F}}{\pi \hbar^2} + \left(\frac{m_{\rm b}}{\hbar^2}\right)^2\frac{\alpha^2+\beta^2}{\pi}~.
\end{equation}

The eigenstates of ${\hat {\cal H}}_0 + {\hat {\cal H}}_{\rm SOC}$ corresponding to the eigenvalues (\ref{eq:bands}) 
are given by the product of a plane wave and a spinor,
\begin{equation}\label{eq:psi}
\Psi_{{\bm k}, \lambda}({\bm r}) = \frac{e^{i {\bm k} \cdot {\bm r}}}{\sqrt{S}} \times \frac{1}{\sqrt{2}} 
\left(
\begin{array}{c}
1 \\
\lambda e^{-i \gamma_{\bm k}}
\end{array}
\right)~,
\end{equation}
where $S$ is the area of the system and $\gamma_{\bm k} = \gamma_{\bm k}(\theta_{\bm k})$ is given by 
\begin{equation}\label{eq:tangamma}
\tan\gamma_{\bm k}=\frac{\alpha \cos{(\theta_{\bm k})} + \beta \sin{(\theta_{\bm k})}}{\beta \cos{(\theta_{\bm k})}
+ \alpha \sin{(\theta_{\bm k})}}~.
\end{equation}
The map 
\begin{equation}
{\bm k} \to {\bm {\hat n}}_{\rm eq}({\bm k}) = (\cos{(\gamma_{\bm k})}, -\sin{(\gamma_{\bm k})})
\end{equation}
between momentum and the unit vector ${\bm {\hat n}}_{\rm eq}$, which parametrizes 
the noninteracting  orientation of the spin texture in momentum space, establishes the helical nature of the model.

Electron-electron interactions in Eq.~(\ref{eq:Hamiltonian}) are described by the usual two-body spin-independent Hamiltonian
\begin{equation}\label{eq:int}
{\hat {\cal H}}_{\rm int} = \frac{1}{2S}
\sum_{{\bm q} \neq 0} \sum_{{\bm k}, {\bm k}'}
\sum_{i,j} v_q 
{\hat \psi}^\dagger_{{\bm k}-{\bm q}, i} 
{\hat \psi}^\dagger_{{\bm k}'+{\bm q}, j} 
{\hat \psi}_{{\bm k}', j} 
{\hat \psi}_{{\bm k}, i}~,
\end{equation}
where $v_q = 2\pi e^2/(\epsilon q)$ is the 2D Fourier transform of the Coulomb interaction ($\epsilon$ being a high-frequency dielectric constant which depends on the specific semiconductor heterojunction in which the 2DEG is created). This specific form of interaction potential applies to a strictly 2D system. The finite width of the quantum well hosting the 2DEG can be easily taken into account by introducing a form factor $F(q)$, which renormalizes the Fourier transform $v_q \to V_q = v_q F(q)$ (see, for example, Ref.~\onlinecite{asgari_prb_2005}). 

As is common in electron-gas theory~\cite{Giuliani_and_Vignale}, the electron density $n$ will be expressed below in terms of 
the more convenient dimensionless Wigner-Seitz parameter $r_s$:
\begin{equation}
r_s = \frac{1}{\sqrt{\pi n a^2_{\rm B}}}~,
\end{equation}
where $a_{\rm B} = \epsilon \hbar^2/(m_{\rm b} e^2)$ is the material Bohr radius.

\subsection{Equations of motion and plasmons}
\label{eq:LRT-EOM}

Collective modes are isolated poles in appropriate dynamical susceptibilities. Plasmons, in particular, are isolated poles in the dynamical density-density response function $\chi_{\rho\rho}(q,\omega)$. 
Note that we are deliberately not denoting the density-density response function by the symbol $\chi_{\rho\rho}({\bm q}, \omega)$, {\it i.e.} we are assuming that the system we are interested in is rotationally invariant and thus its density-density response function depends only on $q =|{\bm q}|$. This happens when $\beta$ {\it or} $\alpha$ is equal to zero. The case $\alpha = \pm \beta$ deserves special attention and will be discussed at a greater length below (see Sect.~\ref{sect:special_comments}). 

In full generality, this response function can be written as
\begin{equation}
\chi_{\rho\rho}(q,\omega) = \frac{{\widetilde \chi}_{\rho\rho}(q,\omega)}{1 -v_q {\widetilde \chi}_{\rho\rho}(q,\omega)}~,
\end{equation}
where ${\widetilde \chi}_{\rho\rho}(q,\omega)$ is the so-called ``proper" density-density response function~\cite{Giuliani_and_Vignale,properdiagrams}, which physically describes the density response to the screened potential. The plasmon mode can be found by solving the equation,
\begin{equation}\label{eq:plasmon_equation}
1 - v_q {\widetilde \chi}_{\rho\rho}(q,\omega) = 0~.
\end{equation}
In this article we are not interested in the dispersion of the plasmon at finite $q$ but only in its limit for $q \to 0$. In this limit we can neglect~\cite{properdiagrams} the distinction between the proper and the full causal response function $\chi_{\rho\rho}(q,\omega)$.  

In Sect.~\ref{sect:proof} we prove that
\begin{equation}\label{eq:calD}
\lim_{\omega \to 0} \lim_{q \to 0} \Re e~\chi_{\rho\rho}(q,\omega) = 
\frac{{\cal D}}{\pi e^2}~\frac{q^2}{\omega^2}~,
\end{equation}
where the quantity ${\cal D}$ depends on density and on SOC. Note the order of limits in Eq.~(\ref{eq:calD}) and everywhere below: the limit $\omega \to 0$ is taken in the 
``dynamical" sense~\cite{Giuliani_and_Vignale}, {\it i.e.} $\omega \gg v_{{\rm F}, \lambda} q$, where $v_{{\rm F}, \lambda}$ is the Fermi velocity for each chiral band.

Before proceeding with the formal proof of Eq.~(\ref{eq:calD}) we highlight its main physical consequences. 
Using Eq.~(\ref{eq:calD}) in Eq.~(\ref{eq:plasmon_equation}) and solving for $\omega$ we find that, to leading order in $q$,
\begin{equation}\label{eq:MainResult}
\omega_{\rm pl}(q\to 0) = \sqrt{\frac{2 {\cal D}}{\epsilon}}~q^{1/2} \equiv \sqrt{\frac{2 \pi n e^2}{m_{\rm pl} \epsilon}}~q^{1/2}~,
\end{equation}
where we have introduced the {\it plasmon mass}
\begin{equation}\label{eq:plasmonmass}
m_{\rm pl} = \frac{\pi n e^2}{\cal D}~.
\end{equation}
The plasmon mass is thus completely controlled by the quantity ${\cal D}$.
In the same limit the imaginary-part of the low-frequency a.c. conductivity $\sigma(\omega) = i e^2 \omega \chi_{\rho\rho}(\omega)/q^2$ has the form
\begin{equation}\label{DrudeWeightForm} 
\Im m~\sigma(\omega) \to \frac{{\cal D}}{\pi \omega}~.
\end{equation}
The a.c. conductivity is a causal response function, which implies that its poles can only lie infinitesimally below the real-frequency axis, {\it i.e.} $\sigma(\omega \to 0) \propto (\omega + i\eta)^{-1}$. It then follows that the real-part of the conductivity has a $\delta$-function Drude peak at $\omega=0$:
\begin{equation}\label{DrudeWeightForm-Real}
\Re e~\sigma(\omega) = {\cal D} \delta(\omega)~.
\end{equation}
The quantity ${\cal D}$ introduced in Eq.~(\ref{eq:calD}) is thus precisely 
the {\it Drude weight}.  In the presence of disorder the $\delta$-function peak in Eq.~(\ref{DrudeWeightForm-Real}) is broadened into a Drude peak, but the Drude weight is preserved.

\subsection{Rigorous definition of the Drude weight}
\label{sect:proof}

We now proceed to demonstrate Eq.~(\ref{eq:calD}) using the ``equations-of-motion" approach. The density operator corresponding to the Hamiltonian (\ref{eq:Hamiltonian}) is given by the usual expression
\begin{equation} \label{eq:density-op}
{\hat \rho}_{\bm q} =\sum_{{\bm k},i} {\hat \psi}^\dagger_{{\bm k}-{\bm q},i} {\hat \psi}_{{\bm k}, i}~,
\end{equation}
and it obeys the standard Heisenberg equation of motion ($\hbar=1$ from now on)
\begin{equation}\label{eq:continuity}
i \partial_t {\hat \rho}_{\bm q} = [{\hat \rho}_{\bm q}, {\hat {\cal H}}] \equiv 
{\bm q} \cdot {\hat {\bm j}}^{({\rm p})}_{\bm q}~,
\end{equation}
which is simply the quantum mechanical version of the continuity equation. Here the so-called 
{\it paramagnetic} current-density operator~\cite{Giuliani_and_Vignale} has the following transparent form:
\begin{equation}\label{eq:current_op-x}
{\hat j}^{({\rm p})}_{{\bm q}, x} = \sum_{{\bm k}, i} 
{\hat \psi}^\dagger_{{\bm k}-{\bm q}, i} \frac{k_x +q_x/2}{m_{\rm b}} {\hat \psi}_{{\bm k}, i} 
+ (\beta {\hat \sigma}^x_{\bm q} -\alpha {\hat \sigma}^y_{\bm q})
\end{equation}
along the ${\hat {\bm x}}$ direction and
\begin{equation}\label{eq:current_op-y}
{\hat j}^{({\rm p})}_{{\bm q}, y} = \sum_{{\bm k}, i} {\hat
\psi}^\dagger_{{\bm k}-{\bm q}, i} 
\frac{k_y +q_y/2}{m_{\rm b}}{\hat \psi}_{{\bm k}, i} 
+ (\alpha {\hat \sigma}^x_{\bm q} -\beta {\hat \sigma}^y_{\bm q})~,
\end{equation}
along the ${\hat {\bm y}}$ direction. In Eqs.~(\ref{eq:current_op-x})-(\ref{eq:current_op-y}) we have introduced the spin-density operators
\begin{equation}\label{eq:spindensity}
{\hat \sigma}^\mu_{\bm q} = \sum_{{\bm k}, i,  j} {\hat \psi}^\dagger_{{\bm k}-{\bm q}, i} \sigma^\mu_{ij}{\hat \psi}_{{\bm k}, j}~.
\end{equation}

We now introduce the causal linear-response functions $\chi_{AB}(\omega)$, which 
are defined by the Kubo ``product"~\cite{Giuliani_and_Vignale},
\begin{eqnarray}
\chi_{AB}(\omega) &=& \frac{1}{S}\langle\langle {\hat A}; {\hat B}\rangle\rangle_\omega
\nonumber\\
& \equiv &-\frac{i}{S} \int_0^{\infty}dt \langle [{\hat A}(t),
{\hat B}(0)]\rangle e^{i\omega t}e^{-\eta t}~,
\end{eqnarray}
where the symbol $\langle {\hat {\cal O}} \rangle$ denotes the expectation value of the operator ${\hat {\cal O}}$ over 
the {\it exact} interacting ground state and $\eta \to 0^+$ is a positive infinitesimal. The dynamical response
function $\langle\langle {\hat A}; {\hat B}\rangle\rangle_\omega$ obeys the
following identity
\begin{equation}\label{eq:eom_general}
\langle\langle {\hat A}; {\hat B}\rangle\rangle_\omega = \frac{1}{\omega}\langle
[{\hat A}, {\hat B}] \rangle +\frac{i}{\omega}
\langle\langle \partial_t {\hat A}; {\hat B}\rangle\rangle_\omega~,
\end{equation}
or,
\begin{equation}\label{eq:eom_general_bis}
\langle\langle {\hat A}; {\hat B}\rangle\rangle_\omega = \frac{1}{\omega}\langle
[{\hat A}, {\hat B}] \rangle -\frac{i}{\omega}
\langle\langle  {\hat A}; \partial_t{\hat B}\rangle\rangle_\omega~.
\end{equation}

Using the continuity equation (\ref{eq:continuity}) and  Eqs.~(\ref{eq:eom_general})-(\ref{eq:eom_general_bis}), the density-density response function $\chi_{\rho\rho}({\bm q},\omega)$ can be expressed in terms of the 
longitudinal paramagnetic current-current response function as,
\begin{eqnarray}
\chi_{\rho\rho}({\bm q},\omega) &\equiv & \frac{1}{S}\langle\langle {\hat \rho}_{\bm q}; {\hat \rho}_{-{\bm q}}\rangle\rangle_\omega \nonumber \\
&=&\frac{1}{S} \frac{1}{\omega}\langle\langle {\bm q} \cdot {\hat {\bm j}}^{({\rm
p})}_{\bm q}; {\hat \rho}_{-{\bm q}}\rangle\rangle_\omega \nonumber\\
&=&\frac{1}{S} \frac{{\bm q}\cdot \langle [{\hat {\bm j}}^{({\rm p})}_{\bm q},
{\hat \rho}_{-{\bm q}}]\rangle}{\omega^2}  + \frac{1}{S}\frac{\langle\langle {\bm q} \cdot {\hat {\bm j}}^{({\rm
p})}_{\bm q};  {\bm q} \cdot {\hat {\bm j}}^{({\rm p})}_{-{\bm
q}}\rangle\rangle_\omega}{\omega^2}~.\nonumber\\
\end{eqnarray}
We remind the reader that in the presence of a vector potential ${\bm A}_{\bm k}$ 
the physical current-density operator ${\hat {\bm j}}_{\bm q}$ is related to the paramagnetic one by
\begin{equation} \label{eq:physicalJ}
{\hat {\bm j}}_{\bm q} ={\hat {\bm j}}^{({\rm p})}_{\bm
q}+\frac{e}{m_{\rm b} c S}\sum_{\bm k}{\bm A}_{{\bm q}-{\bm k}}{\hat \rho}_{\bm k}~.
\end{equation}
The paramagnetic current-current response function 
\begin{equation}
\chi_{j^{({\rm p})}_\ell j^{({\rm p})}_m}({\bm q},\omega) = \frac{1}{S}
\langle\langle {\hat j}^{({\rm p})}_{{\bm q}, \ell}; {\hat j}^{({\rm p})}_{-{\bm
q}, m}\rangle\rangle_\omega
\end{equation}
(here $\ell,m$ label the coordinate indices) is thus related to the physical one by the simple equation~\cite{Giuliani_and_Vignale}
\begin{equation}\label{eq:fundamental}
\chi_{j_\ell j_m}({\bm q},\omega)=
\frac{n}{m_{\rm b}}\delta_{\ell m} + \chi_{j^{({\rm p})}_\ell j^{({\rm
p})}_m}({\bm q},\omega)~.
\end{equation}

For generic values $(\alpha,\beta)$ of the SOC constants, the dynamical response functions of the model described by Eq.~(\ref{eq:Hamiltonian}) are {\it anisotropic}, {\it i.e.} they depend on the direction of ${\bm q}$. However, in the cases of pure Rashba ($\beta = 0$) or pure Dresselhaus ($\alpha = 0$) SOC the ground state of the Hamiltonian ${\hat {\cal H}}_0 + {\hat {\cal H}}_{\rm SOC}$ is rotationally invariant: for the sake of simplicity, in what follows we will restrict our attention to these two ``extreme" cases. From now on we assume $\alpha \neq 0$ and $\beta =0$. In Sect.~\ref{sect:special_comments} we will comment on how our results change in the pure Dresselhaus ($\alpha = 0$ and $\beta \neq 0$) case and in the special case $\alpha = \pm \beta$. Last but not least, we also assume to be in the regime in which both chiral bands are occupied ($\varepsilon_{\rm F} >0$). In this situation the Fermi surface consists of two concentric circles. At low enough densities the topology of the Fermi surface changes dramatically, the occupied states becoming an annulus in momentum space. We will not tackle interaction effects in this interesting but hard to achieve experimentally regime. 

In a homogeneous and isotropic liquid we can decompose the tensor $\chi_{j_\ell j_m}(q,\omega)$ into its longitudinal and transverse components with respect to the direction of ${\bm q}$:
\begin{eqnarray}
\chi_{j_\ell j_m}(q,\omega) &=& \chi_{\rm
L}(q,\omega)\frac{q_\ell q_m}{q^2}\nonumber\\
&+&\chi_{\rm T}(q,\omega)\left(\delta_{\ell m}-\frac{q_\ell
q_m}{q^2}\right)~.
\end{eqnarray}
Using this definition we immediately end up with the following result 
\begin{equation}\label{eq:iso}
\chi_{\rho\rho}(q,\omega) = \frac{1}{S} \frac{{\bm q} \cdot \langle [{\hat {\bm j}}^{({\rm
p})}_{\bm q}, {\hat \rho}_{-{\bm q}}]\rangle}{\omega^2} 
+\frac{q^2}{\omega^2}\left[\chi_{\rm L}(q,\omega) - \frac{n}{m_{\rm b}}\right]~.
\end{equation}
We stress that Eq.~(\ref{eq:iso}) is exact (provided that the ground state is homogenous and isotropic).

The commutator on the r.h.s. of Eq.~(\ref{eq:iso}) can be calculated easily: indeed, the portion of the paramagnetic current operator due to spin-orbit coupling [second terms on the r.h.s. of Eqs.~(\ref{eq:current_op-x})-(\ref{eq:current_op-y})] is proportional to the spin operator 
${\hat {\bm \sigma}}_{\bm q}$ only, which commutes with the density operator ${\hat \rho}_{-{\bm q}}$.
Thus the commutator is found to be equivalent to that of the 2DEG {\it without} any spin orbit coupling. 
It is related to the so-called f-sum rule~\cite{Giuliani_and_Vignale} and is given by
\begin{equation}\label{eq:f-sum-rule}
\frac{1}{S}[{\hat {\bm j}}^{({\rm p})}_{\bm q}, {\hat \rho}_{-{\bm q}}] = {\bm
q}\frac{n}{m_{\rm b}}~.
\end{equation}
Using Eq.~(\ref{eq:f-sum-rule}) into Eq.~(\ref{eq:iso}) we are left with
the following crucially important relation:
\begin{eqnarray}\label{eq:response_parabolic_bis}
\chi_{\rho\rho}(q,\omega) = \frac{q^2}{\omega^2}~\chi_{\rm L}(q,\omega)~.
\end{eqnarray}
Note that the f-sum rule 
is crucial for the cancellation of the diamagnetic $n/m_{\rm b}$ term in the 
square brackets on the r.h.s of Eq.~(\ref{eq:iso}). 

Eq.~(\ref{eq:response_parabolic_bis}) is identical in form with Eq.~(\ref{eq:calD}) provided that we identify ${\cal D}$ with the following dynamical limit:
\begin{equation}\label{eq:Drudeweight}
{\cal D} \equiv \pi e^2 \lim_{\omega \to 0} \lim_{q \to 0} \Re e~\chi_{\rm L}(q,\omega)~.
\end{equation}
This equation is extremely important because it gives us an operational definition of the Drude weight. In order to calculate it we need to compute the dynamical limit of the real part of the longitudinal current-current response function $\chi_{\rm L}(q,\omega)$. Such a microscopic calculation will be carried out below in Sect.~\ref{sect:TDHFtheory} within the so-called time-dependent Hartree-Fock approximation.

\subsection{Broken Galileian invariance}
\label{sect:BGI}

In a standard 2DEG without spin-orbit coupling ($\alpha=\beta=0$) the longitudinal current-current response function obeys the exact relation
\begin{equation}\label{eq:nosoc}
\lim_{\omega \to 0} \lim_{q \to 0} \chi_{\rm L}(q,\omega) = \frac{n}{m_{\rm b}}~,
\end{equation}
a nonperturbative result ({\it i.e.} valid for any strength of electron-electron interactions as long as the 2DEG remains in a translationally-invariant and homogeneous ground state), which is completely independent of complicated exchange and correlation effects. In this case the Drude weight becomes ${\cal D} = \pi n e^2/m_{\rm b}$ and the plasmon mass reduces to the bare electron mass, $m_{\rm pl} = m_{\rm b}$.

The physical reason behind the exact result (\ref{eq:nosoc}) is the following. In the limit $q \to 0$ $\chi_{\rm L}(q,\omega)$ measures the response of the system to a {\it homogeneous} 
time-dependent vector potential ${\bm A}(t)$, {\it i.e.} to a homogeneous electric field ${\bm E}(t)= - c^{-1}d{\bm A}(t)/dt$. In a system with a single parabolic band the usual replacement ${\bm p} \to {\bm p} + e {\bm A}(t)/c$ implies that 
a uniform vector potential couples identically to all the electrons and thus only to the center-of-mass motion. This is immediately seen in first quantization:
\begin{eqnarray}
{\hat {\cal H}_0}({\bm A}) &=& \sum_i \frac{1}{2 m_{\rm b}}\left[{\bm p}_i +\frac{e}{c} {\bm A}(t)\right]^2 \nonumber\\
& =& {\hat {\cal H}_0} + \frac{e}{m_{\rm b} c} {\bm P}_{\rm CM} \cdot {\bm A}(t) +{\cal O}({\bm A}^2)~,
\end{eqnarray}
where ${\bm P}_{\rm CM} = \sum_i {\bm p}_i$ is the centre-of-mass momentum. In the last equality 
terms of order ${\bm A}^2$ have been neglected since we are interested in the linear-response regime. 
Electron-electron interactions are thus completely transparent to ${\bm A}(t)$, since the latter does not probe the relative motion of electrons.

Eq.~(\ref{eq:nosoc}) can be derived by a classical Newton's equation for the centre-of-mass coordinate ${\bm R}_{\rm CM}$:
\begin{equation}\label{eq:Newton}
m_{\rm b} N \frac{d^2 {\bm R}_{\rm CM}}{dt^2}= - e N {\bm E}(t) =  \frac{e}{c} N \frac{d {\bm A}(t)}{dt}~,
\end{equation}
where $N$ is the total number of electrons. Integrating this equation we find 
${\bm V}_{\rm CM}(t) =  [e/(m_{\rm b} c)]~{\bm A}(t)$ or
\begin{equation}
{\bm j}^{({\rm p})}_{{\bm q}={\bm 0}}(t) = n {\bm V}_{\rm CM}(t)= \frac{n}{m_{\rm b}}~\frac{e}{c}~{\bm A}(t)~,
\end{equation}
{\it i.e.} Eq.~(\ref{eq:nosoc}).

To see more formally why Eq.~(\ref{eq:nosoc}) comes about, we can use the exact-eigenstate (Lehmann) representation~\cite{Giuliani_and_Vignale} for the current-current response function:
\begin{eqnarray}\label{eq:lehmann}
\chi_{j_\ell j_m}(q,\omega) &=& 
\frac{n}{m_{\rm b}} \delta_{\ell m} 
+ \frac{1}{S} \sum_n \left(\frac{\langle 0| {\hat j}^{({\rm p})}_{{\bm
q}, \ell}|n\rangle\langle n| {\hat j}^{({\rm p})}_{-{\bm
q}, m}|0\rangle}{\omega-\omega_{n0}+i \eta}\right. \nonumber\\
&-&\left. \frac{\langle 0| {\hat j}^{({\rm p})}_{-{\bm q}, m}|n\rangle\langle n|
{\hat j}^{({\rm p})}_{{\bm q}, \ell}|0\rangle}{\omega+\omega_{n0}+i
\eta}\right)~,
\end{eqnarray}
where the limit $\eta \to 0^+$ is understood. In a translationally invariant system, with or without SOC, the exact eigenstates $|n\rangle$ are eigenstates of the total momentum. In the absence of SOC, moreover, ${\hat j}^{({\rm p})}_{\ell, {\bm q}={\bm 0}}$ coincides with the total momentum [see Eqs.~(\ref{eq:current_op-x})-(\ref{eq:current_op-y})] and thus for $\alpha = \beta =0$ and $q \to 0$ the second term in Eq.~(\ref{eq:lehmann}) vanishes and one is left with Eq.~(\ref{eq:nosoc}). In the presence of SOC, however, ${\hat j}^{({\rm p})}_{\ell, {\bm q}={\bm 0}}$ does not coincide with the total momentum and thus Eq.~(\ref{eq:nosoc}) ceases to be true.

When $\alpha$ (or $\beta$) is non-zero we have
\begin{equation}\label{eq:soc}
\lim_{\omega \to 0} \lim_{q \to 0} \chi_{\rm L}(q,\omega) \neq \frac{n}{m_{\rm b}}~.
\end{equation}
Deviations from the trivial $n/m_{\rm b}$ result are due to {\it both} single- and many-particle effects~\cite{farid_prl_2006}.

The single-particle contribution to the long-wavelength low-energy limit of $\chi_{\rm L}(q,\omega)$ can be found quite easily. 
In Sect.~\ref{sect:interactioncorrections} we will show that if electron-electron interactions are neglected 
\begin{equation}\label{eq:singleparticle}
\lim_{\omega \to 0} \lim_{q \to 0} \chi^{(0)}_{\rm L}(q,\omega) = \frac{n}{m_{\rm b}} - \alpha^2\frac{\nu_0}{2}~,
\end{equation}
where $\nu_0 = m_{\rm b}/\pi$ is the usual 2D parabolic-band density-of-states in the absence of SOC. 
The rest of the paper is mainly devoted to quantifying interaction-corrections to Eq.~(\ref{eq:singleparticle}).

\subsection{Failure of the random phase approximation}
\label{sect:failureRPA}

Before concluding this Section, we would like to emphasize that the popular random phase approximation (RPA) is not capable 
of capturing the subtle renormalizations of the Drude weight due to many-body effects. 

By definition, within RPA the proper density-density response 
function ${\widetilde \chi}_{\rho\rho}(q,\omega)$ is approximated with its noninteracting value~\cite{Giuliani_and_Vignale}:
\begin{equation}\label{eq:RPA}
{\widetilde \chi}_{\rho\rho}(q,\omega) \stackrel{\rm RPA}{\to} \chi^{(0)}_{\rho\rho}(q,\omega) = \frac{q^2}{\omega^2} \chi^{(0)}_{\rm L}(q,\omega)~.
\end{equation}
When Eq.~(\ref{eq:RPA}) is substituted in Eq.~(\ref{eq:Drudeweight}) one finds immediately that the RPA Drude weight is identical to its noninteracting value:
\begin{eqnarray}
{\cal D}_{\rm RPA} &=& \lim_{\omega \to 0} \lim_{q \to 0} \chi^{(0)}_{\rm L}(q,\omega) = \pi e^2\left[\frac{n}{m_{\rm b}} - \alpha^2\frac{\nu_0}{2}\right] \nonumber\\
&\equiv& {\cal D}_0~.
\end{eqnarray}
More physically, the reason why RPA does not capture the subtle interaction renormalizations of ${\cal D}$ is the following. During a plasmon oscillation the Fermi circle oscillates back and forth in momentum space. Due to SOC this oscillatory motion of charge excites spin oscillations. Exchange interactions are of course very sensitive to the spin degrees-of-freedom. The RPA, however, is simply a time-dependent Hartree theory~\cite{Giuliani_and_Vignale}, which treats exactly only the self-consistent {\it electrical} potential,
\begin{equation}
V_{\rm H}({\bm r},t) = \int d^2{\bm r}' \frac{e^2}{\epsilon|{\bm r}- {\bm r}'|}\delta n({\bm r}',t)~, 
\end{equation}
created by the electrons displaced away from the equilibrium position in the presence of the neutralizing background, while completely neglecting the self-consistent exchange field associated with the spin degrees-of-freedom. From this argument it clearly emerges that the minimal theory which can capture interaction-corrections to Eq.~(\ref{eq:singleparticle}) is the time-dependent Hartree-Fock theory.

\section{Microscopic time-dependent Hartree-Fock theory}
\label{sect:TDHFtheory}

In this Section we present a microscopic theory of ${\cal D}$ that takes into account electron-electron interactions in an approximate manner. As discussed in Sect.~\ref{sect:failureRPA}, the minimal approximation that captures the renormalization of ${\cal D}$ due to many-body effects is the so-called time-dependent Hartree-Fock approximation (TDHFA). One of the pleasant properties of the TDHFA is that it is exact to first order in Coulomb interactions. Other advantages, such as its relative simplicity, will be evident below.

As we have amply discussed in the previous Sections, we want to study the response of the system described by the Hamiltonian (\ref{eq:Hamiltonian}) to a weak homogeneous external time-dependent electric field directed along, say, $\bm {\hat x}$. In the gauge in which the scalar potential is zero the electric field is simply described by a time-dependent vector potential: ${\bm E}(t) = - [c^{-1} d A(t)/dt]~{\bm {\hat x}}$. The vector potential enters the Hamiltonian (\ref{eq:Hamiltonian}) {\it via} the usual minimal coupling ${\bm p} \to {\bm p} + e A(t) {\hat {\bm x}}/c$. The parabolic-band part becomes
\begin{eqnarray}\label{eq:Hzerotime}
{\hat {\cal H}}_0(t) &=& \sum_{{\bm k}, i} \frac{\displaystyle \left[k_x + \frac{e}{c}A(t)\right]^2 +k^2_y}{2m_{\rm b}}{\hat \psi}^\dagger_{{\bm k}, i} {\hat \psi}_{{\bm k}, i}~,
\end{eqnarray}
while the SOC part reads
\begin{equation}\label{eq:HSOCtime}
{\hat {\cal H}}_{\rm SOC}(t) = \alpha\sum_{{\bm k}, i, j} 
{\hat \psi}^\dagger_{{\bm k}, i} \left\{\sigma^x_{ij}k_y -\sigma^y_{ij} 
\left[k_x + \frac{e}{c}A(t)\right]\right\}
{\hat \psi}_{{\bm k}, j}~.
\end{equation}
Neglecting terms ${\cal O}(A^2)$, which are beyond linear-response theory, we can write 
the sum of the two terms in Eqs.~(\ref{eq:Hzerotime})-(\ref{eq:HSOCtime}) as
\begin{eqnarray}\label{eq:explicit}
{\hat {\cal H}}_0(t) + {\hat {\cal H}}_{\rm SOC}(t) &=& {\hat {\cal H}}_0 + {\hat {\cal H}}_{\rm SOC} + \frac{e}{m_{\rm b} c} 
P^x_{\rm CM} A(t) \nonumber\\
&-& \alpha \frac{e}{c} {\hat \sigma}^y_{\rm tot} A(t)~.
\end{eqnarray}
Thus, due to SOC, a magneto-electric effect appears~\cite{edelstein_ssc_1990}: a uniform electric field applied along the ${\hat {\bm x}}$ direction acts as a uniform magnetic field in the ${\hat {\bm y}}$ direction [last term in the r.h.s. of Eq.~(\ref{eq:explicit})]. Here ${\hat \sigma}^y_{\rm tot} = {\hat \sigma}^y_{{\bm q} ={\bm 0}}$.

Electron-electron interactions are treated within the Hartree-Fock (HF) mean-field theory in which the two-body term in Eq.~(\ref{eq:int}) 
is approximated as~\cite{Hartree}
\begin{eqnarray}\label{eq:decoupling}
{\hat \psi}^\dagger_{{\bm k}-{\bm q}, i} {\hat \psi}^\dagger_{{\bm k}'+{\bm
q}, j} {\hat \psi}_{{\bm k}', j} {\hat \psi}_{{\bm k}, i} &\approx & 
-:{\hat \psi}^\dagger_{{\bm k}-{\bm q}, i} {\hat \psi}_{{\bm k}', j}: 
\langle {\hat \psi}^\dagger_{{\bm k}'+{\bm q}, j}{\hat \psi}_{{\bm k},
i}\rangle \nonumber \\
&-& :{\hat \psi}^\dagger_{{\bm k}'+{\bm q}, j}{\hat \psi}_{{\bm k}, i}:
\langle {\hat \psi}^\dagger_{{\bm k}-{\bm q}, i}{\hat \psi}_{{\bm k}',
j}\rangle~,\nonumber\\
\end{eqnarray}
where $\langle \dots \rangle$ ($:\dots:$) denote the expectation value over (normal ordering with respect to) the HF ground state~\cite{Giuliani_and_Vignale}. At this point we introduce the spin-density matrix,
\begin{equation}\label{eq:sdm}
\langle{\hat \psi}^\dagger_{{\bm k}, i}{\hat \psi}_{{\bm k}', j}\rangle = \delta_{{\bm k}, {\bm k}'} \rho_{ij}({\bm k})~,
\end{equation}
which just assumes that the mean-field ground state is translationally invariant.  
The interaction contribution to the total Hamiltonian reads
\begin{eqnarray}\label{eq:hf_interaction}
{\hat {\cal H}}_{\rm int} &=& -\frac{1}{S} \sum_{{\bm k}, {\bm k}'}
\sum_{i, j} v_{{\bm k}-{\bm k}'}
\rho_{ji}({\bm k}') :{\hat \psi}^\dagger_{{\bm k}, i}{\hat
\psi}_{{\bm k}, j}:~.
\end{eqnarray}
We parametrize the spin-density matrix $\rho_{ij}({\bm k})$ in a compact form~\cite{mishchenko_prb_2003} 
in terms of the occupation factors, $n_{{\bm k}, \pm}$, of the {\it noninteracting} 
Hamiltonian ${\hat {\cal H}}_0 + {\hat {\cal H}}_{\rm SOC}$ in the eigenstate representation and in the absence of $A(t)$:
\begin{equation}\label{eq:spin-density-matrix}
\rho_{ij}({\bm k}) = \frac{n_{{\bm k}, +} + n_{{\bm k},
-}}{2}\delta_{ij} +\frac{n_{{\bm k}, +} - n_{{\bm k}, -}}{2}~{\hat {\bm n}}({\bm
k})\cdot {\bm \sigma}_{ji}~.
\end{equation}
Here ${\hat {\bm n}}({\bm k})$ is a unit vector 
on the 2D plane which denotes the orientation of the spins in the 
total ``effective" magnetic field.  The idea behind this parametrization is that a	
homogeneous external field (a field with ${\bm q}={\bm 0}$) cannot change
anything but the orientation of the spin, which is encoded in the unit
vector ${\hat {\bm n}} ({\bm k})$. Note that in the absence of the external field ${\hat {\bm n}}({\bm k}) = {\hat {\bm n}}_{\rm eq}({\bm k})$. Eq.~(\ref{eq:spin-density-matrix}) is nevertheless approximate since it assumes the absence of interaction effects in the ground state of the system. 
More explicitly, the Fermi wave vectors are renormalized by electron-electron interactions~\cite{chesi_prb_2007}, $k^{(0)}_{{\rm F}, \pm} \to k_{{\rm F}, \pm}$. 
We will come back to this point below in Sect.~\ref{sect:equilibrium}.

 Using Eq.~(\ref{eq:spin-density-matrix}) in Eq.~(\ref{eq:hf_interaction}), the total mean-field HF Hamiltonian can be written as
\begin{equation} \label{eq:Hamiltonian_HF}
{\hat {\cal H}}_{\rm HF} = \sum_{{\bm k}, i, j} :{\hat
\psi}^\dagger_{{\bm k}, i} 
\left[\delta_{ij}B_0({\bm k}) +{\bm \sigma}_{ij} \cdot {\bm B}({\bm
k}) \right]{\hat \psi}_{{\bm k}, j}:
\end{equation}
where the HF fields are defined by
\begin{equation}\label{eq:B0}
B_0({\bm k}) = \varepsilon(k)  +\frac{e}{m_{\rm b} c}P^x_{\rm CM}A(t)- 
\int \frac{d^2{\bm k}'}{(2\pi)^2} v_{{\bm k}-{\bm k}'} f_{\mathbf + }({\bm k}') 
\end{equation}
and
\begin{equation}\label{eq:def-B}
{\bm B}({\bm k}) = {\bm h}({\bm k}) - \int \frac{d^2{\bm k}'}{(2\pi)^2} v_{{\bm k}-{\bm k}'} f_{-}({\bm k}') {\hat {\bm
n}}({\bm k'})~.  
\end{equation} 
In Eq.~({\ref{eq:def-B}),
\begin{equation}\label{eq:necessary}
{\bm h}({\bm k}) = - \alpha\frac{e}{c}A(t) {\hat {\bm y}} + \alpha k~{\hat {\bm n}}_{\rm eq}({\bm k})
\end{equation}
is an effective magnetic field, which has the external magneto-electric component~\cite{edelstein_ssc_1990} of modulus
\begin{equation}\label{eq:magnetoelectricfield}
B_{\rm ext} =  \alpha \frac{e}{c}A
\end{equation}
arising from the external vector potential and an internal component $\propto  {\hat {\bm n}}_{\rm eq}({\bm k})$, while the last term is the exchange field due to the electron-electron interactions with
\begin{equation}
f_\pm({\bm k}) \equiv \frac{n_{{\bm k}, +} \pm  n_{{\bm k}, -}}{2}
\end{equation}
and
\begin{equation}
{\hat {\bm n}}({\bm k}) \equiv \frac{{\bm B}({\bm k})}{|{\bm B}({\bm k})|}~.
\end{equation}

The noninteracting band-eigenstate occupation factors are given by 
\begin{equation}\label{eq:occupationfactors}
n_{{\bm k}, \pm} = \Theta(\varepsilon_{\rm F} - \varepsilon_\pm({\bm k}))~,
\end{equation}
where $\Theta(x)$ is the standard Heaviside step function. As we have already emphasized above, for pure Rasha SOC ($\beta =0$) 
the momentum occupation factors $n_{{\bm k}, \pm}$ are rotationally-invariant and depend only on $k = |{\bm k}|$ 
(the same is true also for pure Dresselhaus SOC, $\alpha=0$). It is thus extremely convenient to decompose the 
spherically symmetric inter-electron interaction $v_{{\bm k}-{\bm k}'}$ 
in angular momentum components,
\begin{equation}\label{eq:am_expansion}
v_{{\bm k}-{\bm k}'} = \sum_{m=-\infty}^{+\infty}~V_m(k,k')e^{i m (\theta_{\bm
k}-\theta_{{\bm k}'})}~,
\end{equation}
with
\begin{equation}\label{eq:coefficients}
V_m(k,k') = \int_0^{2\pi} \frac{d\theta}{2\pi} e^{-i m \theta} \left.v_q\right|_{q = |{\bm k}-{\bm k}'|}~,
\end{equation}
$\theta$ being the angle between ${\bm k}$ and ${\bm k}'$.

\subsection{Equilibrium HF theory}
\label{sect:equilibrium}

In the absence of the external electric field, {\it i.e.} $A(t)=0$, the unit vector ${\hat {\bm n}}({\bm k})$ coincides with the equilibrium one:
\begin{equation}
{\hat {\bm n}}({\bm k}) \to {\hat {\bm n}}_{\rm eq}({\bm k}) = (\sin{(\theta_{\bm k})}, -\cos{(\theta_{\bm k})})~.
\end{equation}
Substituting Eq.~(\ref{eq:am_expansion}) in Eq.~(\ref{eq:def-B}) and 
performing the angular integration over $\theta_{{\bm k}'}$, we find that the equilibrium solution of Eq.~(\ref{eq:def-B}) reads
\begin{equation} \label{eq:Beq-alpha}
{\bm B}_{\rm eq}({\bm k}) = \left[\alpha k - \int_0^{\infty}\frac{dk'}{2\pi}~k' f_{-}(k') V_1(k,k')\right]{\hat {\bm n}}_{\rm eq}({\bm k})~.
\end{equation}
As expected, in the absence of the electric field, ${\bm B}({\bm k})$ is
oriented along ${\hat {\bm n}}_{\rm eq}({\bm k})$ and it is isotropic. The modulus of ${\bm B}_{\rm eq}({\bm k})$ 
is simply
\begin{equation}\label{eq:equilibriumBvec}
\left|{\bm B}_{\rm eq}({\bm k})\right| = \alpha k + \Sigma_1(k)~,
\end{equation}
and depends only on $k$ with the self-energy $\Sigma_1(k)$ defined by
\begin{equation}\label{eq:selfenergyone}
\Sigma_1(k) = - \int_0^{\infty}\frac{dk'}{2\pi}k'~f_{-}(k') V_1(k,k')~.
\end{equation}
For $\varepsilon_{\rm F} > 0$ the factor $f_-$ in the integrand, being the difference in the occupation of the two bands, picks up contributions 
only from wave vectors $k'$ in the interval $[k_{{\rm F}, +}^{(0)}, k_{{\rm F}, -}^{(0)}]$,  where $k_{{\rm F}, \lambda}^{(0)}$ is given by 
Eq.~(\ref{eq:kF}) with $\beta=0$. The self energy can thus be written as 
\begin{equation}\label{eq:equilibrium_HF_self_energy_numerics}
\Sigma_1(k)=\frac{1}{4\pi}\int_{k_{{\rm F}, +}^{(0)}}^{k_{{\rm F}, -}^{(0)}}dk'~k'V_1(k,k')~.
\end{equation}

In a completely analogous manner, it is possible to find the equilibrium solution of Eq.~(\ref{eq:B0}) which reads  
\begin{equation} \label{eq:B02} 
B_{0, {\rm eq}}({\bm k}) = \varepsilon(k) - \int_0^{\infty}\frac{dk'}{2\pi}~k'f_{+}(k') V_0(k,k')~,
\end{equation}
or,  $B_{0, {\rm eq}}({\bm k}) = \varepsilon(k) +\Sigma_0(k)$ with
\begin{eqnarray}\label{eq:selfenergyzero}
\Sigma_0(k) &=& - \frac{1}{2\pi}\int_{0}^{k_{{\rm F}, +}^{(0)}}dk'~k' V_0(k,k') \nonumber\\
&-& \frac{1}{4\pi}\int_{k_{{\rm F}, +}^{(0)}}^{k_{{\rm F}, -}^{(0)}}dk'~k' V_0(k,k')~. 
\end{eqnarray}

Finally, the complete HF bands are given by
\begin{equation} \label{eq:HFbands}
E_{{\rm HF}, \lambda}(k)= \varepsilon(k)  +\Sigma_0(k) + \lambda [\alpha k + \Sigma_1(k)]~,
\end{equation}
and the quasiparticle effective mass $m^\star_\lambda$ for the $\lambda$-th band 
can be defined as
\begin{equation}\label{eq:mass0}
\frac{k_{{\rm F}, \lambda}^{(0)}}{m_{\lambda}^\star} \equiv \left.\frac{\partial E_{{\rm HF}, \lambda}(k)}{\partial k}
\right|_{k=k_{{\rm F}, \lambda}^{(0)}}~.
\end{equation}

Rigorously speaking, the HF bands and the corresponding Fermi wave vectors ($k_{{\rm F}, \pm}$) should be calculated in a fully self-consistent manner. We have done this and we find that the repopulation of the energy bands in the ground state due to interactions is a very small effect. In fact, the difference between $k^{(0)}_{{\rm F}, \pm}$ and $k_{{\rm F}, \pm}$ 
is less than $0.5$\% over the entire range of parameters we have considered. We have thus ignored this small effect throughout this article and used $k^{(0)}_{{\rm F}, \pm}$ in all calculations.

\subsection{The non-equilibrium problem: linearization of the HF equation}
\label{sect:TDHFA}
We now proceed to solve Eq.~(\ref{eq:def-B}) in the presence of the external electric field 
by linearizing it around the equilibrium solution ${\bm B}_{\rm eq}({\bm k})$. To this end we write
\begin{equation}\label{eq:Bnext}
{\bm B}({\bm k}) = {\bm B}_{\rm eq}({\bm k}) + \delta {\bm B}({\bm k})
\end{equation}
and
\begin{equation}\label{eq:delta_n_linearized}
{\hat {\bm n}}({\bm k}) = {\hat {\bm n}}_{\rm eq}({\bm k})+ \frac{\delta {\bm B}_\perp({\bm k})}{|{\bm B}_{\rm eq}({\bm k})|} + {\cal O}((\delta {\bm B})^2)~,
\end{equation}
where $\delta {\bm B}_\perp({\bm k}) = \delta {\bm B}({\bm k}) - {\hat {\bm n}}_{\rm eq} ({\bm k})[{\hat {\bm n}}_{\rm eq} ({\bm k})\cdot \delta {\bm B}({\bm k})]$ is the component of $\delta {\bm B}({\bm k})$ perpendicular to ${\hat {\bm n}}_{\rm eq} ({\bm k})$. 
We now make the following {\it Ansatz} for $\delta {\bm B}({\bm k})$:
\begin{eqnarray} \label{eq:def_B_linearized}
\delta {\bm B}({\bm k}) &=& [\delta B_{{\rm L},1}(k)\cos{(\theta_{\bm k})}] {\hat {\bm n}}_{\rm eq} ({\bm k}) 
\nonumber \\ &-& [\delta B_{{\rm T}, 1}(k) \sin(\theta_{\bm k})] {\hat {\bm z}}\times {\hat {\bm n}}_{\rm eq} ({\bm k})~.
\end{eqnarray}
We note that the {\it Ansatz} has to be consistent with the underlying model Hamiltonian. Indeed, for the pure Rashba model, 
${\hat {\bm n}}_{\rm eq}({\hat {\bm k}}) = {\hat {\bm k}} \times {\hat {\bm z}}$ and ${\hat {\bm z}} \times {\hat {\bm n}}_{\rm eq}({\hat {\bm k}}) = {\hat {\bm k}}$.  
Since the magneto-electric field is in the ${\hat {\bm y}}$-direction [see Eq.~(\ref{eq:necessary})] its components along ${\hat {\bm z}} \times {\hat {\bm n}}_{\rm eq}$ and ${\hat {\bm n}}_{\rm eq}$ 
are proportional to $\sin(\theta_{\bm k})$ and $\cos(\theta_{\bm k})$, respectively. This justifies the particular form of Eq.~(\ref{eq:def_B_linearized}). Using Eq.~(\ref{eq:def_B_linearized}) in Eq.~(\ref{eq:delta_n_linearized}) we find that
\begin{equation}\label{eq:def_n_linearized}
\delta {\hat {\bm n}} ({\bm k}) \equiv  {\hat {\bm n}}({\bm k}) - {\hat {\bm n}}_{\rm eq} ({\bm k}) = - \frac{\delta B_{{\rm T},1}(k)\sin(\theta_{\bm k})}{|{\bm B}_{\rm eq}({\bm k})|}~{\hat {\bm k}}~.
\end{equation}

Substituting Eqs.~(\ref{eq:def_B_linearized}) 
and~(\ref{eq:def_n_linearized}) in Eqs.~(\ref{eq:def-B}), integrating over $\theta_{{\bm k}'}$, 
and keeping only terms that are linear in $\delta {\bm B}({\bm k})$, we find
\begin{eqnarray}
\frac{1}{2} [\delta B_{{\rm L},1}(k) &+& \delta B_{{\rm T},1}(k)] =  B_{\rm ext}\nonumber \\
&-&\int_0^{\infty}\frac{dk'}{4\pi}~k' f_{\mathbf -}(k') V_0(k,k')\frac{\delta
B_{{\rm T},1}(k')}{|{\bm B}_{\rm eq}({\bm k}')|}~,\nonumber\\
\end{eqnarray}
and
\begin{eqnarray} 
\frac{1}{2}[\delta B_{{\rm L},1}(k) &-& \delta B_{{\rm T}, 1}(k)]=  
\int_0^{\infty}\frac{dk'}{4\pi} k' f_-(k') V_2(k,k')\nonumber\\
&\times&\frac{\delta B_{{\rm T}, 1}(k')}{|{\bm B}_{\rm eq}({\bm k}')|}~.
\end{eqnarray}

Summing and subtracting these two equations we finally find the following integral equation for the transverse $\delta B_{{\rm T},1}(k)$ component:
\begin{equation}\label{eq:linear_rashba}
\delta B_{{\rm T},1}(k) = B_{\rm ext} - \int_0^{\infty}dk'~{\cal K}_{\rm T}(k,k') \delta B_{{\rm T},1}(k') 
\end{equation}
where the kernel ${\cal K}_{\rm T}(k,k')$ is defined by
\begin{eqnarray}
{\cal K}_{\rm T}(k,k') &=& \frac{1}{4\pi}k' f_-(k') \frac{V_0(k,k')+V_2(k,k')}{|{\bm B}_{\rm eq}({\bm k}')|} \nonumber \\
&=& \frac{1}{4\pi}k' f_-(k') \frac{V_0(k,k')+V_2(k,k')}{\alpha k' + \Sigma_1(k')}~.
\end{eqnarray}
Once Eq.~(\ref{eq:linear_rashba}) has been solved self-consistently for $\delta B_{{\rm T}, 1}(k)$, the longitudinal component $\delta B_{{\rm L},1}(k)$ can be calculated from
\begin{equation}
\delta B_{{\rm L},1}(k) = B_{\rm ext} - \int_0^{\infty}dk'~{\cal K}_{\rm L}(k,k')~\delta B_{{\rm T},1}(k')
\end{equation}
with
\begin{equation}
{\cal K}_{\rm L}(k,k') = \frac{1}{4\pi}k' f_{\mathbf -}(k') \frac{V_0(k,k')-V_2(k,k')}{\alpha k' + \Sigma_1(k')}~.
\end{equation}

For future reference, it is very convenient to rewrite Eq.~(\ref{eq:linear_rashba}) 
in a dimensionless form. To this end, we scale all the wave vectors with the 2DEG Fermi wave vector 
$k_{\rm F} = \sqrt{2 \pi n}$ in the absence of SOC, all energies with $\varepsilon_{{\rm F},0} = k^2_{\rm F}/(2 m_{\rm b})$, 
the pseudopotentials $V_m$ with $2\pi e^2/(\epsilon k_{\rm F})$, and, finally, we introduce the dimensionless SOC constant~\cite{comment_upper_bound} ${\bar \alpha} =m_{\rm b} \alpha/k_{\rm F}$ and the dimensionless quantity $u =\delta B_{{\rm T}, 1}/B_{\rm ext}$. From now on, symbols with a bar over them denote dimensionless quantities. In these units Eq.~(\ref{eq:linear_rashba}) reads
\begin{equation}\label{eq:integral_equation_u}
u(x) = 1 + \frac{r_s}{2\sqrt{2}}\int_{\Lambda_+}^{\Lambda_-}dx'~x'
\frac{{\bar V}_0(x,x')+{\bar V}_2(x,x')}{2 {\bar \alpha} x' +  {\bar \Sigma}_1(x')}~u(x')~.
\end{equation}
Here $x =k/k_{\rm F}$, $x' = k'/k_{\rm F}$, 
\begin{equation}\label{eq:lambdas}
\Lambda_\pm = \frac{k_{{\rm F}, \pm}^{(0)}}{k_{\rm F}} = \mp {\bar \alpha} + \sqrt{1 - {\bar \alpha}^2}~,
\end{equation}
and ${\bar \Sigma}_1(x)$ is the dimensionless version of the self-energy introduced in Eq.~(\ref{eq:equilibrium_HF_self_energy_numerics}):
\begin{equation}
{\bar \Sigma}_1(x) = \frac{\Sigma_1}{\varepsilon_{{\rm F}, 0}} = \frac{r_s}{\sqrt{2}} F(x)
\end{equation}
with
\begin{equation}
F(x) = \int_{\Lambda_+}^{\Lambda_-}dx'~x'{\bar V}_1(x,x')~.
\end{equation}
\subsection{Interaction corrections to the Drude weight and renormalization of the in-plane spin susceptibility}
\label{sect:interactioncorrections}

We are now in the position to evaluate the interaction corrections to the Drude weight from the definition in Eq.~(\ref{eq:Drudeweight}). 

We need to evaluate the longitudinal response $\chi_{\rm L}$ to a uniform vector potential $A$ in the $\omega \to 0$ limit. 
We thus have to evaluate the change in longitudinal physical current due to a uniform electric field applied, say, 
along the ${\hat {\bm x}}$ direction:
\begin{equation}
\delta j_x = \lim_{A \to 0} 
\left[\langle {\hat j}_{{\bm q} = {\bm 0} ,x}\rangle_A - \langle {\hat j}_{{\bm q} = {\bm 0} ,x}\rangle_{A=0}\right]~,
\end{equation}
where the physical current operator ${\hat {\bm j}}_{\bm q}$ has been introduced in Eq.~(\ref{eq:physicalJ}). 
Recalling the definition of the spin-density matrix $\rho_{ij}({\bm k})$ we find that
\begin{equation}\label{eq:delta_j_x_rashba} 
\delta j_x = \frac{n}{m_{\rm b}}\frac{eA}{c} + \frac{1}{S}\sum_{{\bm k},i}\frac{k_x}{m_{\rm b}}\delta \rho_{ii}({\bm k}) - \alpha \frac{1}{S} \sum_{{\bm k}, i, j}\sigma^y_{ij}\delta \rho_{ij}({\bm k})~.
\end{equation}
Since the diagonal components of the spin-density matrix do not change under the application of a uniform magnetic field we get the following important relation
\begin{equation}
\delta j_x = \frac{n}{m_{\rm b}}\frac{eA}{c} - \alpha~\delta \sigma^{y}~,
\end{equation}
where 
\begin{eqnarray}\label{eq:delta_sigma_y}
\delta \sigma^{y}&=& \frac{1}{S} \sum_{{\bm k}, i, j} \sigma^y_{ij} ~\delta \rho_{ij}({\bm k}) 
= 2 \int \frac{d^2{\bm k}}{(2\pi)^2} f_-(k)~\delta {\hat n}_y
\nonumber \\
&=& -2\int \frac{d^2{\bm k}}{(2\pi)^2} f_-(k) \frac{\delta B_{{\rm T},1}(k)}{|{\bm B}_{\rm eq}({\bm k})|} ~\sin^2{(\theta_{\bm k})}~. 
\end{eqnarray}
Performing the angular integration we finally find that 
\begin{equation}
\delta \sigma^{y} = \chi_{\sigma^y \sigma^y} B_{\rm ext}~,
\end{equation}
where the magneto-electric field $B_{\rm ext}$ has been introduced above in Eq.~(\ref{eq:magnetoelectricfield}) and where the in-plane spin susceptibility is given by
\begin{eqnarray}\label{eq:spin-susceptibility}
\chi_{\sigma^y \sigma^y} &=& \frac{1}{B_{\rm ext}}\int_{k_{{\rm F}, +}^{(0)}}^{k_{{\rm F}, -}^{(0)}}\frac{d k}{4\pi}~k 
\frac{\delta B_{{\rm T}, 1}(k)} {|{\bm B}_{\rm eq}({\bm k})|}\nonumber \\ 
&=& \frac{\nu_0}{2}\int_{\Lambda_+}^{\Lambda_-}d x~\frac{x u(x)} { 2{ \bar \alpha}x + {\bar \Sigma}_1(x)}~.
\end{eqnarray} 
In the noninteracting $r_s \to 0$ limit the vertex correction $u$ tends to unity and the self-energy $\Sigma_1$ to zero: in this limit Eq.~(\ref{eq:spin-susceptibility}) reproduces the well known result for the in-plane spin susceptibility of a 2DEG with Rashba SOC, {\it i.e.} $\chi^{(0)}_{\sigma^y \sigma^y}= \nu_0/2$.

Finally, using Eq.~(\ref{eq:Drudeweight}) we find that the Drude weight is given by
\begin{equation}\label{eq:connectionDrudespin}
{\cal D} = \pi e^2 \frac{\delta j_x}{e A/c} = \pi e^2\left(\frac{n}{m_{\rm b}} - \alpha^2 \chi_{\sigma^y \sigma^y}\right)~.
\end{equation}
This is the most important result of this work. It states that the corrections due to SOC and many-body effects 
to the universal $\pi e^2 n/m_{\rm b}$ Drude weight of a standard parabolic-band 2DEG are completely controlled by the uniform in-plane spin susceptibility $\chi_{\sigma^y \sigma^y}$ in the dynamical limit. Note that, even though the particular derivation we have given in this Section seems to be related to (and thus dependent on) the TDHFA,  Eq.~(\ref{eq:connectionDrudespin}) is exact and stems directly from Eq.~(\ref{eq:current_op-x}).

In the noninteracting limit $\chi_{\sigma^y \sigma^y} \to \nu_0/2$ and thus, in the same limit,
\begin{equation} 
{\cal D}_0 = \pi e^2\left( \frac{n}{m_{\rm b}} - \alpha^2 \frac{\nu_0}{2}\right)~.
\end{equation}
In Sect.~\ref{sect:numerics} we will present numerical results for the ratio ${\cal D}/{\cal D}_0$ as evaluated from the HF expression for the in-plane spin susceptibility in Eq.~(\ref{eq:spin-susceptibility}). Normally electron-electron interactions enhance the spin susceptibility: we thus anticipate that the Drude weight of the interacting system is {\it smaller} than its value ${\cal D}_0$ in the absence of interactions.

Before concluding this Section we derive a semi-analytical expression for $\chi_{\sigma^y \sigma^y}$ up
to first order in the coupling constant $e^2$. To this order of perturbation theory 
the solution of Eq.~(\ref{eq:integral_equation_u}) can be found analytically with the result
\begin{equation}\label{eq:perturbative}
u(x) = 1 + \frac{r_s}{2\sqrt{2} \bar{\alpha}}g(x)
\end{equation}
where
\begin{equation}
g(x) = \int_{\Lambda_+}^{\Lambda_-}dx' \frac{{\bar V}_0(x,x')+{\bar
V}_2(x,x')}{2}~.
\end{equation}
We notice that the perturbative solution (\ref{eq:perturbative}) is not of the first order in $r_s$, since ${\bar \alpha} = m_{\rm b} \alpha/k_{\rm F}$ and $\Lambda_\pm$ also depend on density 
{\it via} the Fermi wave vector. In the presence of SOC, interaction effects are not solely controlled by $r_s$. Substituting Eq.~(\ref{eq:perturbative}) in Eq.~(\ref{eq:spin-susceptibility}) and expanding the ratio in the integrand of this equation in powers of $e^2$ up to first order we finally find that
\begin{equation}\label{eq:linear-in-r_s}
\frac{\chi_{\sigma^y \sigma^y}}{\chi^{(0)}_{\sigma^y \sigma^y}} = 1+ \frac{r_s {\cal A}}{4\sqrt{2} {\bar{\alpha}}^2}~,
\end{equation}
with
\begin{equation}\label{eq:calA}
{\cal A} = \int_{\Lambda_+}^{\Lambda_-}dx~\left[g(x)-\frac{F(x)}{x}\right]~.
\end{equation}
A plot of ${\cal A}$ as a function of $r_s$ for different values of ${\bar \alpha}$ is reported in Fig.~\ref{fig:one}: 
we clearly see that ${\cal A}$ is positive and that thus the in-plane spin susceptibility is enhanced by electron-electron interactions (at least to first order in $e^2$).
\begin{figure}[t]
\centering
\includegraphics[width=0.98\linewidth]{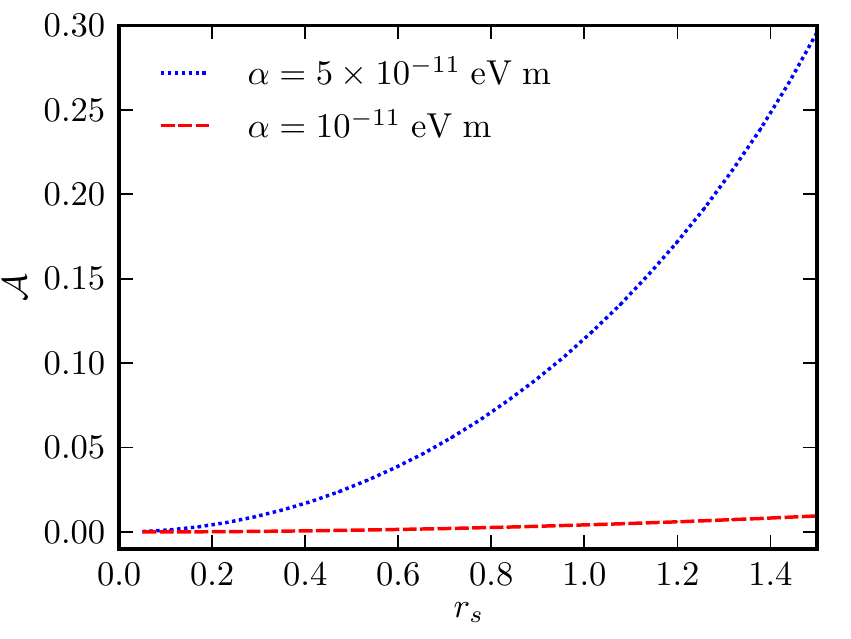} 
\caption{(Color online) The quantity ${\cal A}$ in Eq.~(\ref{eq:calA}) as a function of $r_s$ for two values of the Rashba SOC strength $\alpha$. 
Notice that ${\cal A} >0$ and thus $\chi_{\sigma^y \sigma^y}/\chi^{(0)}_{\sigma^y \sigma^y} > 1$.\label{fig:one}}  
\end{figure}
In the high-density and/or weak-SOC limit ($m_{\rm b} \alpha \ll k_{\rm F}$ or, equivalently, ${\bar \alpha} \ll 1$) we can approximate ${\cal A}$ in the following manner:
\begin{eqnarray}\label{eq:calAweak}
{\cal A} & \to & (\Lambda_- -\Lambda_+) \left[g(x)-\frac{F(x)}{x}\right]_{x=1} \nonumber\\
&=& 2{\bar \alpha}[g(1)-F(1)]~.
\end{eqnarray}
In the same limit
\begin{eqnarray}
g(1) - F(1) &\to& 2{\bar \alpha} \left[ \frac{{\bar V}_0(1,1)+{\bar V}_2(1,1)}{2} - {\bar V}_1(1,1)\right] \nonumber\\
&= & \frac{4 {\bar \alpha}}{3\pi}~,
\end{eqnarray}
the last equality being valid only for unscreened Coulomb interactions [see Eq.~(\ref{eq:analytical_coefficients}) below]. In this case and for ${\bar \alpha} \to 0$ we find ${\cal A} \to 8 {\bar \alpha}^2/(3\pi)$. Using this result in Eq.~(\ref{eq:linear-in-r_s}) we find a rigorous result for the spin susceptibility enhancement to linear order in $r_s$:
\begin{equation}\label{eq:spinweakSOC}
\frac{\chi_{\sigma^y \sigma^y}}{\chi^{(0)}_{\sigma^y \sigma^y}} \to 1+ \frac{\sqrt{2}}{3\pi}r_s~.
\end{equation}

We finally remark that, in the oversimplified case of ultra-short-range interactions, 
\begin{equation}\label{eq:ultrashortrange}
v_q = {\rm constant} = \frac{2 \pi e^2}{\epsilon \kappa}~,
\end{equation}
Eqs.~(\ref{eq:integral_equation_u}) and~(\ref{eq:spin-susceptibility}) can be solved analytically. In this case indeed all the moments $V_m(k,k')$ of the inter-particle interaction but the $m=0$ one are zero.
The solution of the integral equation (\ref{eq:integral_equation_u}) is a constant $u = [1- (2 \kappa a_{\rm B})^{-1}]^{-1}$ and the in-plane spin susceptibility turns out to be equal to $u$:
\begin{equation}\label{eq:chi-simple}
\frac{\chi_{\sigma^y \sigma^y}}{\chi^{(0)}_{\sigma^y \sigma^y}} = u = \frac{1}{1- (2 \kappa a_{\rm B})^{-1}}>1~.
\end{equation}
\subsection{Interaction corrections to the optical spin Hall conductivity}

It turns out that the in-plane spin susceptibility $\chi_{\sigma^y \sigma^y}$ introduced in the previous Section controls also the ``optical spin Hall conductivity" $\sigma_{\rm SH}(\omega)$ of the Rashba model. This was first shown by Dimitrova~\cite{dimitrova_prb_2005}. For the sake of completeness, we briefly summarize here the key steps of the derivation.

The spin operator ${\hat \sigma}^y_{\bm q}$ at ${\bm q} = {\bm 0}$, ${\hat \sigma}^y_{\rm tot}$, satisfies a simple equation of motion:
\begin{equation} \label{eq:partial-sigmay}
 i \partial_t {\hat \sigma}^y_{\rm tot} = [{\hat \sigma}^y_{\rm tot}, {\hat {\cal H}}] = [{\hat \sigma}^y_{\rm tot}, {\hat {\cal H}}_{\rm SOC}] = 
 - 4 i m_{\rm b} \alpha {\hat j}^z_y~,
\end{equation}
where the ${\bm q}={\bm 0}$ ${\hat {\bm z}}$-spin current operator in the ${\hat {\bm y}}$ direction, ${\hat j}^z_y$, is defined by (see for example Ref.~\onlinecite{sinova_prl_2004})
\begin{equation}
{\hat j}^z_y = \frac{1}{2}\sum_{{\bm k}, i, j} \frac{k_y}{m_{\rm b}}{\hat \psi}^\dagger_{{\bm k}, i}\sigma^z_{ij}{\hat \psi}_{{\bm k}, j}~.
\end{equation}
The spin Hall conductivity $\sigma_{\rm SH} (\omega)$ describes a ${\hat {\bm z}}$-polarized spin current flowing in the ${\hat {\bm y}}$ direction in response to a homogeneous (${\bm q} = {\bm 0}$) electric field ${\bm E} = E_x {\hat {\bm x}}$ along the ${\hat {\bm x}}$ direction:
\begin{equation}\label{eq:definitionsigmash}
j^z_y = \sigma_{\rm SH} E_x~.
\end{equation}
From Eq.~(\ref{eq:partial-sigmay}) it is immediately evident that in the d.c. limit $\omega\tau \to 0$ ($\tau$ is the electron-impurity scattering time) the spin Hall conductivity is zero since in a steady state $\langle \partial_t {\hat \sigma}^{y}_{\rm tot} \rangle = 0$.  This is the limit that is relevant to d.c. transport.  The vanishing of the {\it transport} spin Hall conductivity in the Rashba model has been widely discussed in the literature (see {\it e.g.} Ref.~\onlinecite{raimondi_prb_2005}). 

Here we are interested in the high-frequency or clean limit, $\omega \tau \to \infty$, which can in principle be probed in time-resolved experiments with photo-excited carriers~\cite{ruzicka_prb_2008} or in ballistic transport. In this limit the following analysis is particularly useful. 
We first notice from Eq.~(\ref{eq:definitionsigmash}) that the optical spin Hall conductivity is related to the spin-current response function by 
\begin{eqnarray}\label{eq:SHconductivity}
\sigma_{\rm SH}(\omega) &=& \frac{i e}{\omega} \chi_{j^z_y j_x^{\rm (p)}} (\omega) = \frac{i e}{\omega} \langle\langle {\hat j}^z_y; {\hat j}_x^{\rm (p)} \rangle\rangle_\omega~,
\end{eqnarray}
where ${\hat j}_x^{\rm (p)} = {\hat j}^{\rm (p)}_{{\bm q} = {\bm 0}, x}$ is the ${\hat {\bm x}}$ component of the ${\bm q} = {\bm 0}$ paramagnetic current operator in Eq.~(\ref{eq:current_op-x}). We then substitute 
Eq.~(\ref{eq:partial-sigmay}) in the response function on the r.h.s. of Eq. (\ref{eq:SHconductivity}) and we use Eq.~(\ref{eq:eom_general}). We get
\begin{eqnarray}\label{eq:chain}
\langle\langle {\hat j}^z_y; {\hat j}_x^{\rm (p)} \rangle\rangle_\omega &=& - \frac{1}{4 m_{\rm b} \alpha}\langle\langle \partial_t {\hat \sigma}^y_{\rm tot}; {\hat j}_x^{\rm (p)} \rangle\rangle_\omega \nonumber \\
& = & - \frac{1}{4 m_{\rm b} \alpha}\Big\{-i\omega \Big[\langle\langle {\hat \sigma}^y_{\rm tot}; {\hat j}_x^{\rm (p)} \rangle\rangle_\omega \nonumber\\
&-& \frac{1}{\omega}\langle [{\hat \sigma}^y_{\rm tot}, {\hat j}_x^{\rm (p)}]\rangle \Big]\Big\} \nonumber\\
&=& -\frac{i\omega}{4 m_{\rm b}} \langle\langle {\hat \sigma}^y_{\rm tot}; {\hat \sigma}^y_{\rm tot} \rangle\rangle_\omega~,
\end{eqnarray}
where we have used that $\langle\langle {\hat \sigma}^y_{\rm tot}; P^x_{\rm CM}\rangle\rangle_\omega =0$ and that $[{\hat \sigma}^y_{\rm tot}, P^x_{\rm CM}] =0$. The former is a consequence of the fact that total momentum is a conserved quantity (in the absence of impurities) even in the presence of Rashba SOC, while the latter is a trivial commutation rule. Using Eq.~(\ref{eq:chain}) in Eq.~(\ref{eq:SHconductivity}) we finally find
\begin{equation}\label{eq:SpinHall-chi}
\sigma_{\rm SH} (\omega)= \frac{e}{4 m_{\rm b}} \chi_{\sigma^{y} \sigma^{y}} (\omega)~. 
\end{equation}
In the high-frequency or clean $\omega \tau \to \infty$ limit and for noninteracting electrons we have 
$\chi_{\sigma^{y} \sigma^{y}} \to \nu_0/2$ and thus Eq.~(\ref{eq:SpinHall-chi}) gives the well-known ``universal" value $\sigma_{\rm SH}(\omega\tau \to \infty) = e/(8 \pi)$. As we have seen above, however, electron-electron interactions enhance the high-frequency spin susceptibility, thus yielding an enhancement of the optical spin Hall conductivity. Using Eq.~(\ref{eq:spinweakSOC}) we immediately find that for $\omega\tau \to \infty$
\begin{equation}\label{eq:SpinHalllinear-rs}
\sigma_{\rm SH} = \frac{e}{8\pi}\left(1+ \frac{\sqrt{2}}{3\pi}r_s\right) > \frac{e}{8\pi}~.
\end{equation}
Eq.~(\ref{eq:SpinHalllinear-rs}) has to be compared with Eq.~(36) in Ref.~\onlinecite{dimitrova_prb_2005} where an identical result was found modulo the sign of the second term in round brackets. Dimitrova indeed predicts a {\it suppression}~\cite{dimitrova_prb_2005} of the spin Hall conductivity due to interactions rather than an enhancement.

For the case of ultra-short-range interactions [see Eq.~(\ref{eq:ultrashortrange})] we find that
\begin{equation}
\sigma_{\rm SH} = \frac{e}{8\pi} \left( 1- \frac{1}{2\kappa a_{\rm B}} \right)^{-1} > \frac{e}{8\pi}~.
\end{equation}  

Before concluding this Section we would like to mention that it is possible to derive a relation similar to that in Eq.~(\ref{eq:SpinHall-chi}) 
for the spin Galvanic effect~\cite{ganichev_nature_2002}, {\it i.e.} the generation of a charge current in the ${\hat {\bm x}}$ direction in response to a homogeneous Zeeman magnetic field ${\bm B} = B_y {\hat {\bm y}}$ applied along the ${\hat {\bm y}}$ direction:
\begin{equation}\label{eq:definitionsigmasg}
j_x = \sigma_{\rm SG} B_y~.
\end{equation}
Following a procedure analogous to the one that led to Eq.~(\ref{eq:SpinHall-chi}), we find
\begin{equation}\label{eq:SpinGalvanicEffect-chi}
\sigma_{\rm SG} (\omega) = \alpha \frac{g \mu_{\rm B}}{2} \chi_{\sigma^{y} \sigma^{y}} (\omega)~,
\end{equation}
where $g$ is the material Land\'e gyromagnetic factor and $\mu_{\rm B}$ is the Bohr magneton.

\subsection{Interaction-induced enhancement of the Rashba SOC}

From the functional form (\ref{eq:HFbands}) of the HF bands of the Rashba model it is evident that, for a given density, the real part of the a.c. conductivity 
\begin{equation}
\Re e~\sigma(\omega) = - e^2 \lim_{q \to 0}  \frac{\omega}{q^{2}} \Im m~\chi_{\rho \rho}(q,\omega)~,
\end{equation} 
is finite ({\it i.e.} absorption occurs) only in a finite interval of frequencies:  $\Delta_{+} < \omega < \Delta_{-}$, where
\begin{equation}
\left\{
\begin{array}{l} 
\Delta_+ =  E_{{\rm HF}, +}(k_{{\rm F}, +}^{(0)}) - E_{{\rm HF}, -}(k_{{\rm F}, +}^{(0)}) \vspace{0.1 cm}\\ 
\Delta_-  = E_{{\rm HF}, +}(k_{{\rm F}, -}^{(0)}) - E_{{\rm HF}, -}(k_{{\rm F}, -}^{(0)})
\end{array}
\right.~.
\end{equation}
In the noninteracting limit these bounds are~\cite{magarill_jept_2001,mishchenko_prb_2003}:  
$\Delta^{(0)}_+ = 2 \alpha k_{{\rm F}, +}^{(0)}$ and $\Delta^{(0)}_- =2 \alpha k_{{\rm F}, -}^{(0)}$. 
In the interacting case we can define 
$\Delta_+ \equiv 2 {\widetilde \alpha}_+ k_{{\rm F}, +}^{(0)} $ and 
$\Delta_- \equiv 2 {\widetilde \alpha}_- k_{{\rm F}, -}^{(0)}$, where
\begin{equation}\label{eq:ratiosSOCint}
\left\{
\begin{array}{l}
{\displaystyle \frac{{\widetilde \alpha}_+}{\alpha} = 1 + \frac{r_s \Lambda_{-}}{2 \sqrt{2} {\bar \alpha}} \int_{\Lambda_+}^{\Lambda_-}dx'~x' {\bar V}_1 (\Lambda_+, x')} \vspace{0.2 cm}\\
{\displaystyle \frac{{\widetilde \alpha}_-}{\alpha}  = 1 + \frac{r_s \Lambda_{+}}{2 \sqrt{2} {\bar \alpha}}   \int_{\Lambda_+}^{\Lambda_-}dx'~x'{\bar V}_1 (\Lambda_-, x')} 
\end{array}
\right.~.
\end{equation}
We can thus view ${\widetilde \alpha}_\pm$ as effective Rashba SOC strengths renormalized by electron-electron 
interactions~\cite{chen_prb_1999,shekhter_prb_2005}. The dependencies of ${\widetilde \alpha}_\pm$ on $r_s$ and $\alpha$ will be illustrated below in Sect.~\ref{sect:numerics}.

In high-density and/or weak SOC limit (${\bar \alpha} \ll  1$) we find that 
$\Delta_+ = \Delta_- = 2 {\widetilde \alpha} k_{\rm F}$ with (restoring physical dimensions 
for a moment to make contact with earlier work)
\begin{eqnarray}\label{eq:chenraikh}
\frac{\widetilde \alpha}{\alpha} = 1+ \frac{m_{\rm b}}{2\pi \hbar^{2}} \int_{0}^{2 \pi} \frac{d\theta}{2 \pi}~\cos{(\theta)}~\left.v_{q}\right|_{q =2 k_{\rm F} \sin{(\theta/2)}}~,
\end{eqnarray}
in perfect agreement with the work by Chen and Raikh~\cite{chen_prb_1999}. 
\subsection{The pure Dresselhaus case and the $\alpha = \pm \beta$ case}
\label{sect:special_comments}

Before turning to the numerical results, we would like to mention that all our results apply equally well to the pure Dresselhaus model, {\it i.e.} for $\alpha = 0$ and finite $\beta$. 
Indeed, replacing the Rashba interaction by the Dresselhaus interaction has the only effect of changing the phase $\gamma_{\bm k}$ of the eigenspinors  $\Psi_{{\bm k}, \lambda}({\bm r})$ 
in Eq.~(\ref{eq:psi}) by $\pi/2$, leaving the band energies in Eq.~(\ref{eq:bands}) unchanged. We have also checked that the statement at the beginning of this Section is true by applying the TDHFA to the pure Dresselhaus model.

The model with $\alpha = \pm \beta$ is much more subtle. For $\alpha = \beta$, for example, 
${\hat {\cal H}}_{\rm SOC}$ reads
\begin{eqnarray}\label{eq:alphaequalbeta}
{\hat {\cal H}}_{\rm SOC} &=& \alpha \sum_{{\bm k}, i, j} {\hat \psi}^\dagger_{{\bm k}, i} (k_x + k_y)(\sigma^x_{ij} -\sigma^y_{ij})
{\hat \psi}_{{\bm k}, j} \nonumber\\
& = & 2 \alpha \sum_{{\bm k}, i, j} {\hat \psi}^\dagger_{{\bm k}, i} \sigma^-_{ij} k_+{\hat \psi}_{{\bm k}, j} ~,
\end{eqnarray}
where $\sigma^-_{ij} \equiv (\sigma^x_{ij} -\sigma^y_{ij})/\sqrt{2}$ and $k_+ \equiv (k_x + k_y)/\sqrt{2}$. From the second line in Eq.~(\ref{eq:alphaequalbeta}) 
we immediately see that the magneto-electric field generated by the application of a uniform vector potential ${\bm A}(t)$ is 
\begin{equation}
{\bm B}_{\rm ext} = \alpha\frac{e}{c} (A_x + A_y) (1,-1)~,
\end{equation}
{\it i.e.} it is parallel to $\left.{\hat {\bm n}}_{\rm eq}\right|_{\alpha=\beta} = (1,-1)/\sqrt{2}$, 
and does not affect the spin orientation. Thus, in the case $\alpha = \pm \beta$ 
a uniform vector potential does {\it not} reorient spins. The plasmon mass and the Drude weight are thus completely unrenormalized by electron-electron interactions. Of course the plasmon dispersion at finite $q$ will be sensitive to interactions.

\section{Numerical results}
\label{sect:numerics}

We now turn to a presentation of our main numerical results. As far as the material parameters are concerned, in this article we present results for a 2DEG hosted in a InAs quantum well. In this material the bare electron mass is $m_{\rm b} \approx 0.023~m_{\rm e}$, where $m_{\rm e}$ is the electron mass in vacuum, and the high-frequency dielectric constant is $\epsilon \approx 15$. The material Bohr radius turns out to be 
$a_{\rm B} \approx 348$~\AA. As a consequence, a Wigner-Seitz density parameter $r_s =1$ corresponds to a rather low electron density, $n \approx 2.6 \times 10^{10}~ {\rm cm}^{-2}$. The SOC strength in InAs varies in the range~\cite{nitta_prl_1997} $\alpha \approx (1-6) \times 10^{-11}~{\rm eV}~{\rm m}$.

For the numerical calculations we have used a model interaction potential of the form
\begin{equation}\label{eq:potential}
v_q = \frac{2\pi e^2}{\epsilon ( q +\xi q_{\rm TF})}~,
\end{equation}
where $q_{\rm TF}=2/a_{\rm B}$ is the 2D Thomas-Fermi screening wave vector in the absence of SOC and $\xi \in [0,1]$ is a dimensionless control parameter; $\xi = 0$ implies unscreened Coulomb interactions while $\xi =1$ implies Thomas-Fermi screened 
Coulomb interactions. The TDHFA is well known to overestimate many-body effects when the unscreened Coulomb potential is used. (As we have already mentioned earlier, when the unscreened Coulomb potential is used the TDHFA is exact to first order in $e^2$.) On the other hand, when statically screened Thomas-Fermi interactions are used many-body effects are typically largely underestimated. Thus the spirit of the control parameter $\xi$ is to provide us with upper and lower bounds for the strength of interaction corrections to the various observables we present in this Section.

For $\xi=0$ the coefficients $V_m(k,k')$ of the angular-momentum expansion in Eq.~(\ref{eq:am_expansion}) can be calculated
analytically: in dimensionless units these are given by
\begin{eqnarray}\label{eq:analytical_coefficients}
{\bar V}_m(x,x') &=& \int_0^{\infty}dt~J_{m}(tx)J_{m}(tx') \nonumber \\
&=&\frac{x'^{m}}{x^{m+1}}\frac{\Gamma(m+1/2)}{\Gamma(m+1)\Gamma(1/2)}\nonumber\\
&\times&~_2F_1(m+1/2,1/2,m+1,x'^2/x^2)~,\nonumber\\
\end{eqnarray}
for $x>x'$. Here $J_m(z)$, $\Gamma(z)$, and $_2F_1(a,b,c,z)$ are the Bessel function, the Euler Gamma function, and the hypergeometric function, respectively. 
For $x<x'$ one needs to interchange $x \leftrightarrow x'$ in Eq.~(\ref{eq:analytical_coefficients}). For $\xi \neq 0$ the pseudopotentials ${\bar V}_m(x,x')$ have to be calculated numerically.

Fig.~\ref{fig:two} shows the HF bands $E_{{\rm HF}, \pm}(k)$ in Eq.~(\ref{eq:HFbands}) for unscreened Coulomb interactions, while Fig.~\ref{fig:three} illustrates the HF self-energies $\Sigma_0(k)$ and $\Sigma_1(k)$, defined in Eqs.~(\ref{eq:selfenergyzero}) and~(\ref{eq:equilibrium_HF_self_energy_numerics}), respectively. Note that $\Sigma_0(k)$ is negative while $\Sigma_1(k)$ is positive.

\begin{figure}
\centering
\includegraphics[width=1.0\linewidth]{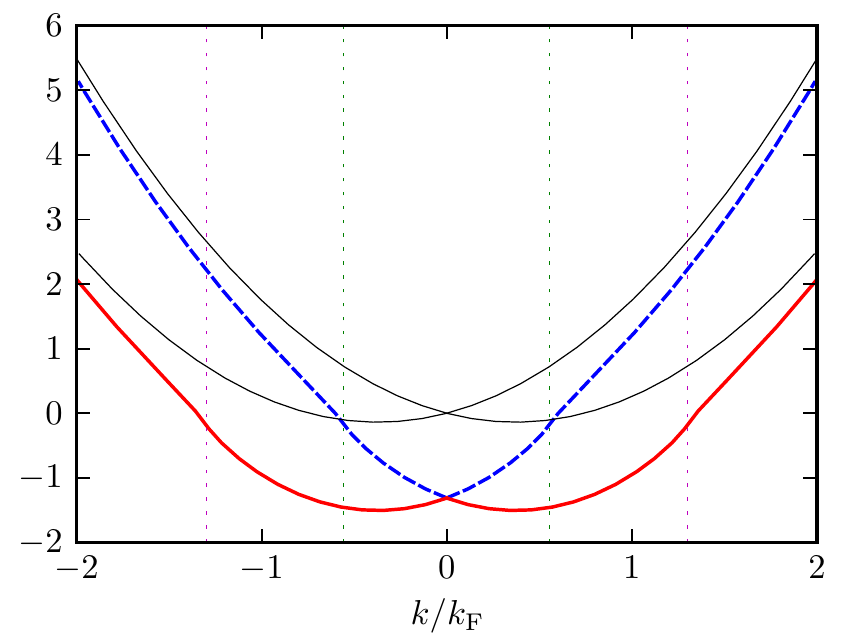} 
\caption{(Color online) The renormalized Hartree-Fock energy bands $E_{{\rm HF}, \pm}(k)$ (in units of $\varepsilon_{{\rm F}, 0}$) as functions 
of $k$ (in units of $k_{\rm F}$) for $r_s =1$ and $\alpha= 5 \times 10^{-11}~{\rm eV}~{\rm m}$ for the case of unscreened ($\xi =0$) Coulomb interactions. The dashed (blue) line denotes the minority band [$E_{{\rm HF}, +}(k)$] while the solid
(red) line denotes the majority band [$E_{{\rm HF}, -}(k)$]. The thin black lines are the energy bands for the noninteracting case. 
The vertical lines denote the location of $\pm k_{{\rm F}, \pm}^{(0)}$. \label{fig:two}}
\end{figure}
%
\begin{figure}
\centering
\includegraphics[width=1.0\linewidth]{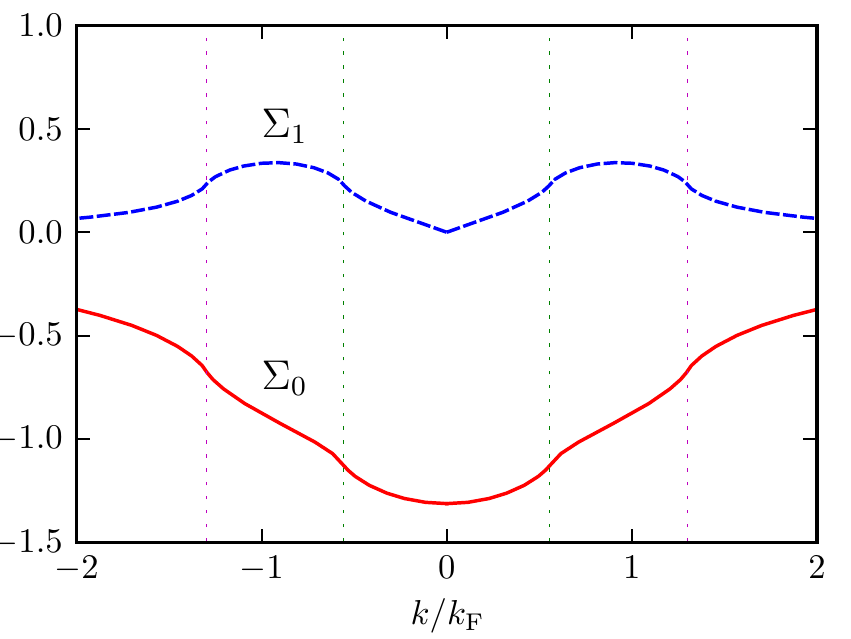} 
\caption{(Color online) The Hartree-Fock self energies $\Sigma_0(k)$ (solid line) and $\Sigma_1(k)$ (dashed line) (in units of $\varepsilon_{{\rm F}, 0}$) as functions of momentum $k$ (in units of $k_{\rm F}$) for $r_s=1$ and $\alpha=5 \times 10^{-11}~{\rm eV}~{\rm m}$. The vertical lines denote the location of $\pm k_{{\rm F}, \pm}^{(0)}$.\label{fig:three}}
\end{figure}

In Figs.~\ref{fig:four}a) and b) we present the minority $m^\star_+$ and majority $m^\star_-$ effective masses 
as functions of $r_{\rm s}$ for Thomas-Fermi screened interactions, as calculated from Eq.~(\ref{eq:mass0}). (It is very well know that, to avoid artifacts of the HF theory, it is necessary to screen the Coulomb interaction to get meaningful results for the quasiparticle effective mass. The derivative of the HF quasiparticle energy indeed diverges at the Fermi surface and hence the quasiparticle effective mass vanishes.) In the case of no SOC ($\alpha=0$) one finds that the quasiparticle effective mass is suppressed by electron-electron interactions. This result stems from exchange interactions and is the dominant effect at weak coupling~\cite{asgari_prb_2005}, {\it i.e.} for $r_s \lesssim 1$. The suppression of the quasiparticle effective mass for $\alpha=0$ shown in Figs.~\ref{fig:four}a)-\ref{fig:four}b) extends to larger values of $r_s$ because of the strong value of the screening parameter ($\xi =1$) we have used. We notice that the impact of SOC is {\it opposite} in different bands 
[this is ultimately due to the dependence of the factor $k^{(0)}_{{\rm F}, \lambda}$ in the l.h.s. of Eq.~(\ref{eq:mass0}) on $\alpha$]: 
as we can see in Fig.~\ref{fig:four}a), SOC further suppresses the quasiparticle effective mass in the minority band. 
On the other hand, as shown in Fig.~\ref{fig:four}b), SOC enhances the quasiparticle effective mass in the majority band. 

In the inset to Fig.~\ref{fig:four}b) we plot the quantities 
$\Delta v^\star_\lambda \equiv v^\star_\lambda - \left. v^\star_\lambda \right|_{\alpha = 0}$ as functions of ${\bar \alpha}$ 
and for a fixed value of $r_s$ ($r_s = 0.25$), where the quasiparticle velocities 
$v^\star_\lambda$ are defined by
\begin{equation}\label{eq:velocities}
v^\star_\lambda = \left.\frac{\partial E_{{\rm HF}, \lambda}(k)}{\partial k}\right|_{k=k_{{\rm F}, \lambda}^{(0)}}~.
\end{equation}
Differently from the effective mass results, we see that $\Delta v^\star_\lambda$ is practically the same for both $\lambda = \pm$
bands and that corrections linear in ${\bar\alpha}$ are absent, in agreement with Refs.~\onlinecite{saraga_prb_2005,chesi_arxiv_2010}.

\begin{figure*}
\centering
\includegraphics[width=0.48\linewidth]{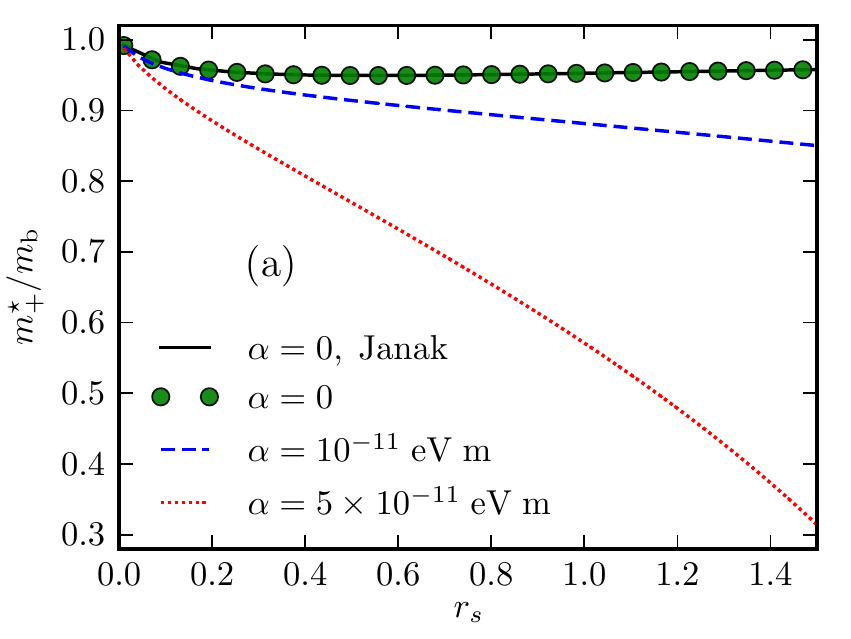}
\hspace{.5cm}
\includegraphics[width=0.48\linewidth]{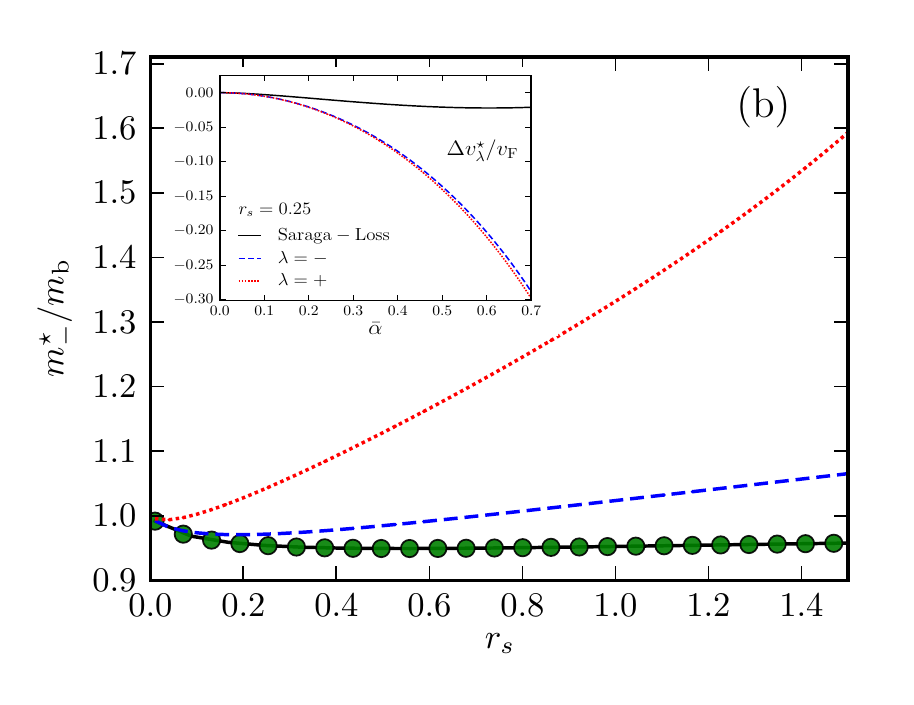} 
\caption{(Color online) The minority [panel a)] and majority [panel b)] 
quasiparticle effective masses $m^\star_\pm$ (in units of the bare mass $m_{\rm b}$)  
as functions of $r_{\rm s}$ for different values of $\alpha$. These results have been obtained by using the definition (\ref{eq:mass0}) and fully-screened Thomas-Fermi interactions ($\xi =1$). The solid line represents the classic result by Janak~\cite{janak_pr_1969}: our results for $\alpha = 0$ (filled circles) are in excellent agreement with the analytical expression (16) in Ref.~\onlinecite{janak_pr_1969}. 
The inset to panel b) illustrates $\Delta v^\star_\pm \equiv v^\star_\pm - \left. v^\star_\pm \right|_{\alpha = 0}$ (in units of $v_{\rm F} \equiv k_{\rm F}/m_{\rm b}$) as functions of ${\bar \alpha}$ 
and for $r_s = 0.25$ (dashed and dotted lines). The solid line is the weak-SOC analytical result (74) in Ref.~\onlinecite{saraga_prb_2005}. Notice that our numerical results extend up to a large value of the
SOC constant since ${\bar\alpha} =0.7$ corresponds to $\alpha \sim  38 \times 10^{-11}~{\rm eV}~{\rm m}$.\label{fig:four}} 
\end{figure*}

In Fig.~\ref{fig:five} we report the solution $\delta B_{{\rm T}, 1}(k)$ of the integral equation (\ref{eq:linear_rashba}) for both unscreened and screened interactions. It is important to note that $\delta B_{{\rm T}, 1}(k)$ in units of the bare effective magnetic field $B_{\rm ext} = eA\alpha/(\hbar c)$ is larger than unity. Kinks are seen in $\delta B_{{\rm T}, 1}(k)$ at $k = k_{{\rm F}, \lambda}^{(0)}$, which are especially visible at $\xi=0$. We also clearly see how the amplitude of $\delta B_{{\rm T}, 1}(k)$ decreases with increasing $\xi$.

\begin{figure*}
\centering
\includegraphics[width=0.48\linewidth]{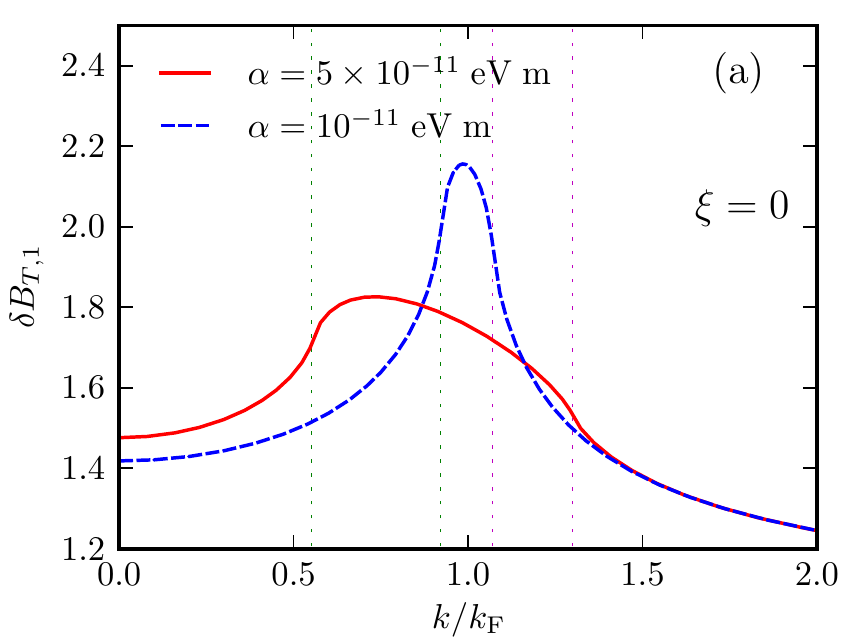} 
\hspace{0.5cm}
\includegraphics[width=0.48\linewidth]{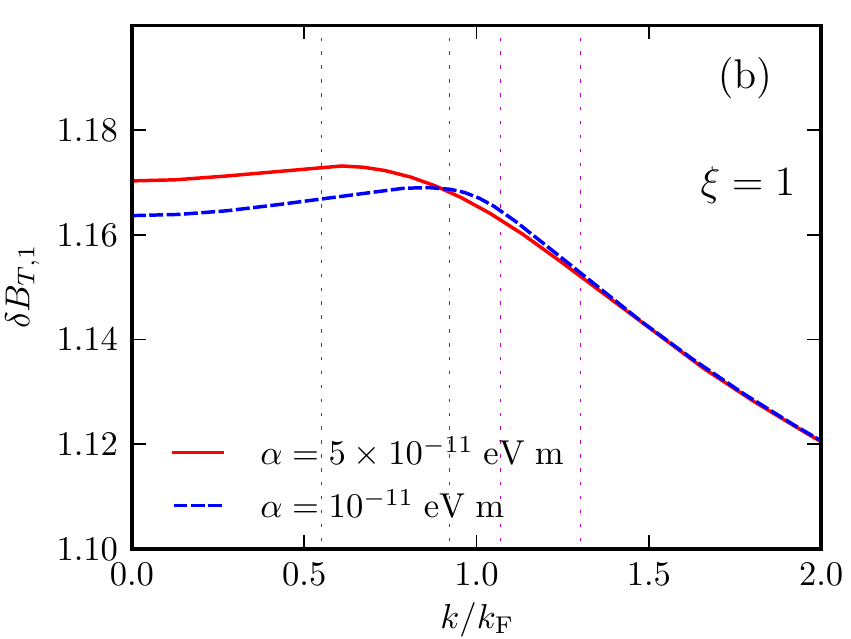} 
\caption{(Color online) The vertex correction $\delta B_{{\rm T}, 1}(k)$ [in units of $B_{\rm ext} = eA\alpha/(\hbar c)$] 
as a function of $k$ (in units of $k_{\rm F}$) for $r_{\rm s}=1$ and different values of $\alpha$. 
Panel a) Results for unscreened Coulomb interactions ($\xi=0$). Panel b) Results for fully-screened Thomas-Fermi interactions ($\xi=1$). Note that when expressing $\delta B_{{\rm T}, 1}(k)$ in units of $B_{\rm ext}$ a factor $\alpha$ is extracted. 
Also note that for $\alpha = 5 \times 10^{-11}~{\rm eV}~{\rm m}$, $k_{{\rm F}, +}^{(0)}/k_{\rm F} \simeq 0.6$ and $k_{{\rm F}, -}^{(0)}/k_{\rm F} \simeq 1.3$. For $\alpha = 10^{-11}~{\rm eV}~{\rm m}$, $k_{{\rm F}, +}^{(0)}/k_{\rm F} \simeq 0.9$ and  $k_{{\rm F}, -}^{(0)}/k_{\rm F} = 1.1$.\label{fig:five}}
\end{figure*}

A plot of the in-plane spin susceptibility $\chi_{\sigma^{y}\sigma^{y}}$ in units of the noninteracting value $\chi^{(0)}_{\sigma^{y}\sigma^{y}} = \nu_0/2$ is presented in Fig.~\ref{fig:six}. As expected, the in-plane spin susceptibility is {\it enhanced} by electron-electron interactions. 
Overscreening their strength by setting $\xi =1$ in Eq.~(\ref{eq:potential}) substantially reduces the enhancement~\cite{yarlagadda_prb_1989}. Note also that increasing the SOC strength $\alpha$ the ratio $\chi_{\sigma^{y}\sigma^{y}}/\chi^{(0)}_{\sigma^{y}\sigma^{y}}$ increases.

\begin{figure*}
\centering
\includegraphics[width=0.48\linewidth]{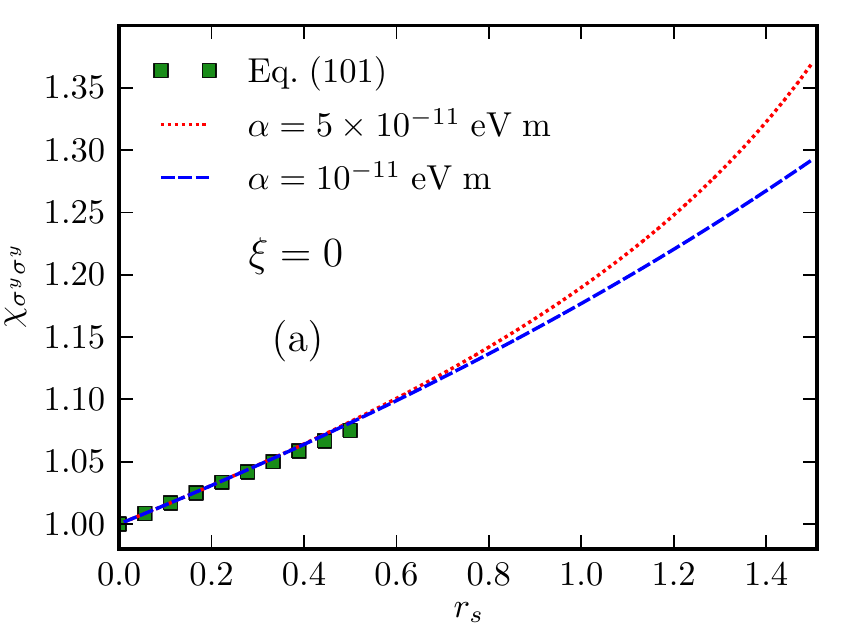}
\hspace{0.5cm}
\includegraphics[width=0.48\linewidth]{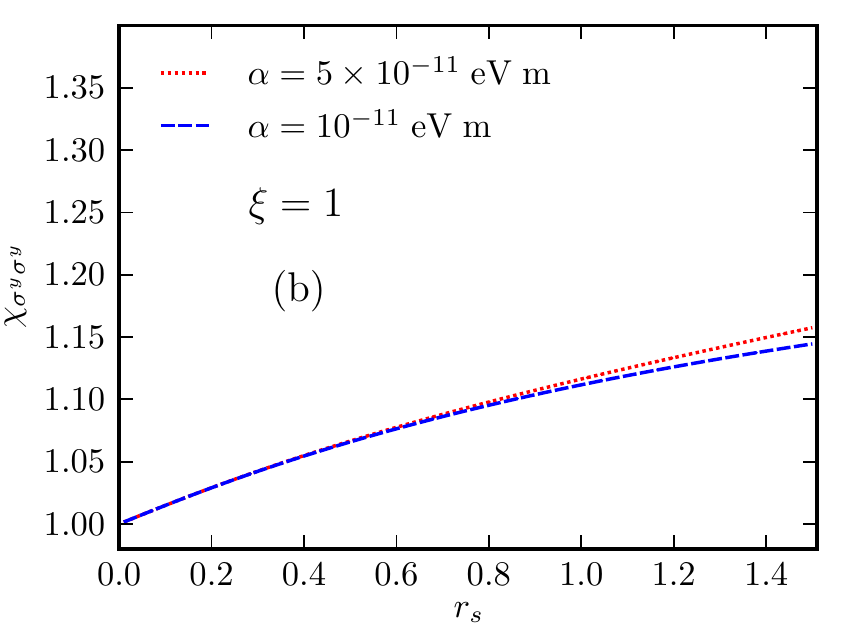}
\caption{(Color online) The in-plane spin susceptibility $\chi_{\sigma^y\sigma^y}$ (in units of the noninteracting value, $\chi^{(0)}_{\sigma^y\sigma^y} = \nu_0/2$) as a function of $r_s$ and for different values of $\alpha$. Panel a) Results for unscreened Coulomb interactions ($\xi=0$). Panel b) Results for fully-screened Thomas-Fermi interactions ($\xi=1$).\label{fig:six}}
\end{figure*}

Fig.~\ref{fig:seven} shows the most important result of this work, {\it i.e.} the 
renormalization of the Drude weight ${\cal D}$ due to interactions. There we indeed plot the ratio 
between ${\cal D}$ and its noninteracting value ${\cal D}_0$. Since the spin susceptibility is enhanced by interactions, 
${\cal D}$ is {\it suppressed}. The suppression is quite large within truly first-order perturbation theory ($\xi=0$) and increases with increasing $\alpha$. The plasmon mass is thus enhanced by the interactions and thus the plasmon frequency is reduced by the combined effect of SOC and interactions with respect to the standard frequency of plasmons in the absence of SOC.

\begin{figure*}
\centering
\includegraphics[width=0.48\linewidth]{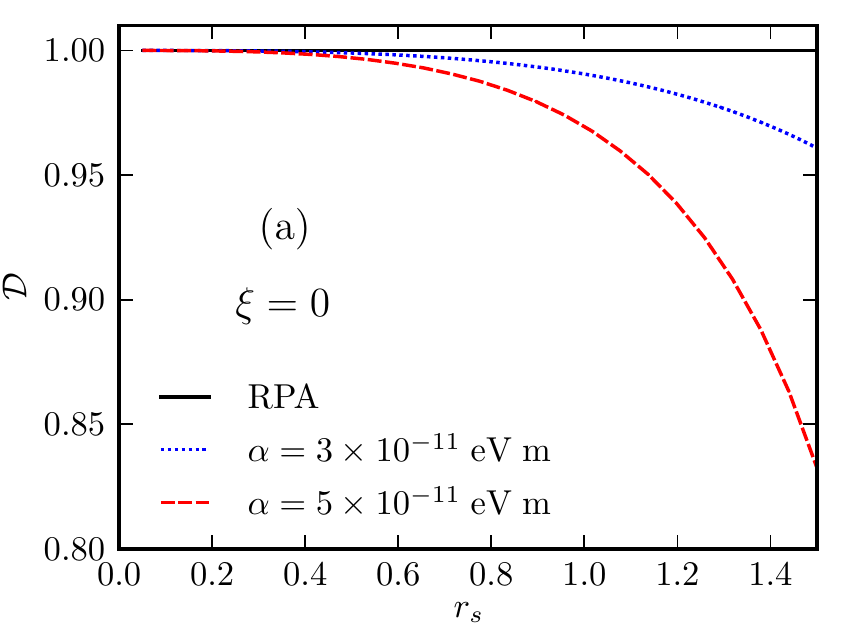} 
\hspace{0.5cm}
\includegraphics[width=0.48\linewidth]{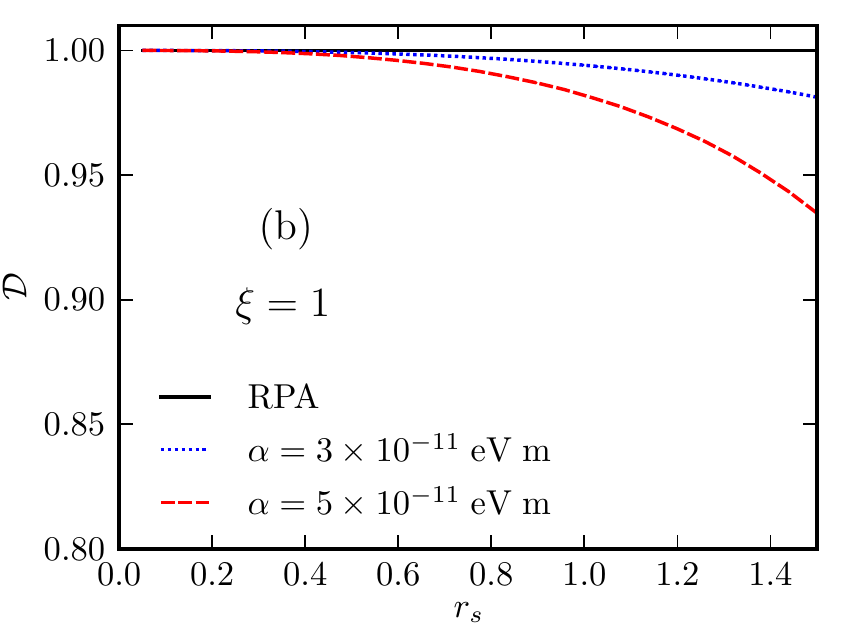} 
\caption{(Color online) The Drude weight ${\cal D}$ [in units of the noninteracting value, ${\cal D}_0 = \pi e^2(n/m_{\rm b} - \alpha^{2} \nu_{0}/2)$], calculated from Eq.~(\ref{eq:connectionDrudespin}), is plotted as a function of $r_s$ and for different values of $\alpha$. Panel a) Results for unscreened Coulomb interactions ($\xi=0$). Panel b) Results for fully-screened Thomas-Fermi interactions ($\xi=1$). 
\label{fig:seven}}
\end{figure*}

The enhancement of SOC due to interactions is illustrated in Figs.~\ref{fig:eight}-\ref{fig:nine} where we have plotted ${\widetilde \alpha}_\pm/\alpha$ as calculated from Eq.~(\ref{eq:ratiosSOCint}). These two figures refer to two different values of $\xi$. For the sake of comparison, in Fig.~\ref{fig:nine} we have also plotted the weak-SOC result by Chen and Raikh~\cite{chen_prb_1999} [see Eq.~(\ref{eq:chenraikh})]. From Fig.~\ref{fig:eight} we see that the enhancement of SOC is pretty large for unscreened Coulomb interactions and that it {\it decreases} for increasing SOC strength.

\begin{figure*}
\centering
\includegraphics[width=0.48\linewidth]{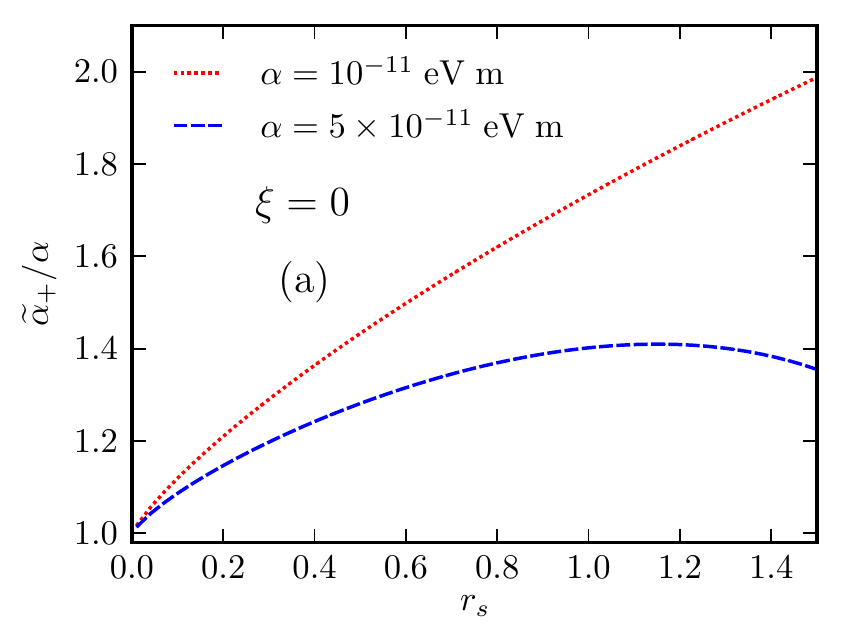}
\hspace{0.5cm}
\includegraphics[width=0.48\linewidth]{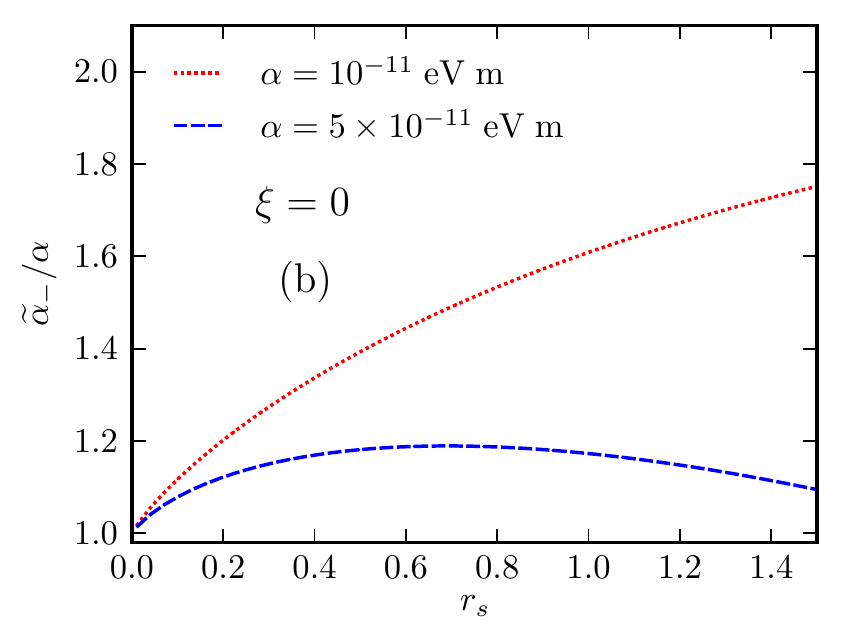}
\caption{(Color online) Renormalized SOC coupling strengths ${\widetilde \alpha}_\pm$ (in units of the bare value $\alpha$) 
for the minority ($+$) and majority ($-$) bands as functions of $r_s$ and for different values of $\alpha$. These results refer to 
unscreened Coulomb interactions ($\xi = 0$). \label{fig:eight}}
\end{figure*}
%

\begin{figure*}
\centering
\includegraphics[width=0.48\linewidth]{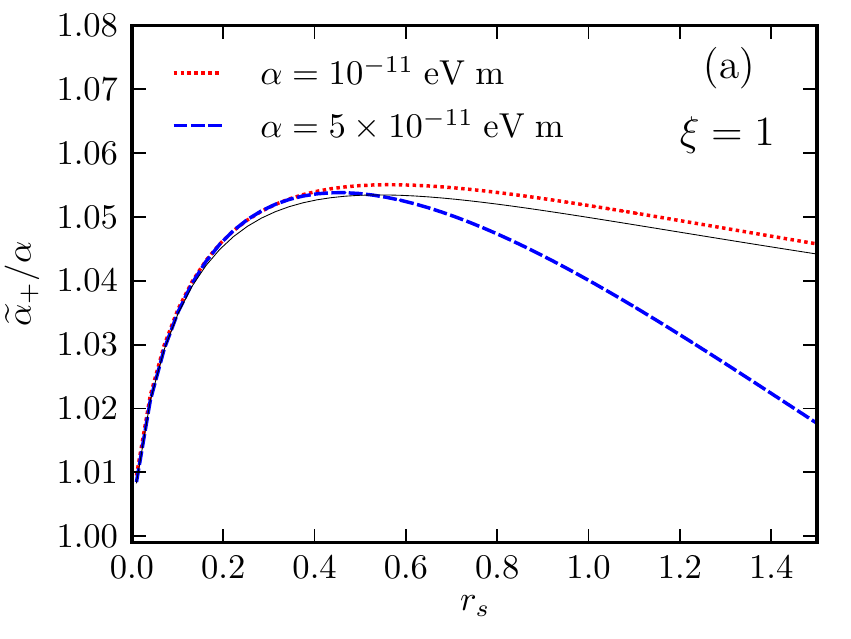} 
\hspace{0.5cm}
\includegraphics[width=0.48\linewidth]{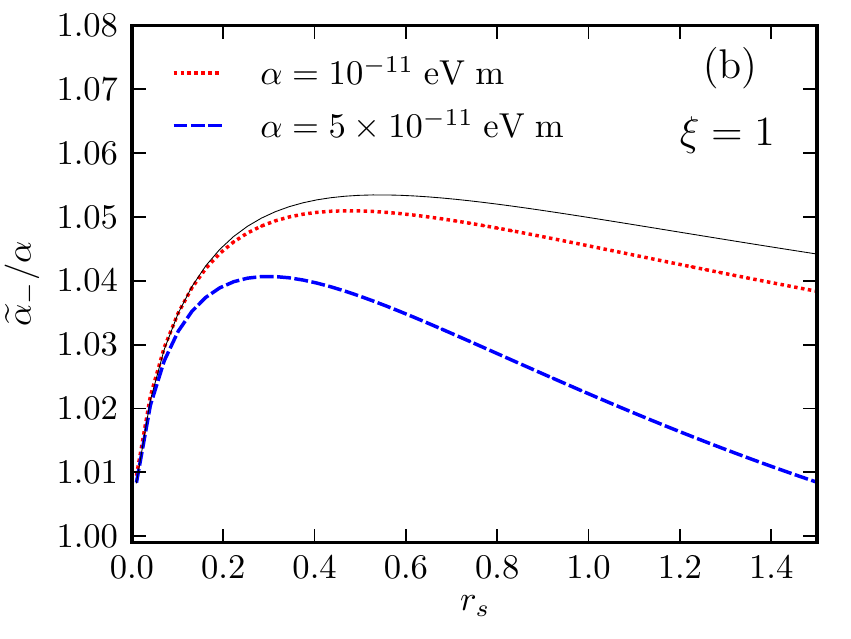}
\caption{(Color online) Same as in Fig.~\ref{fig:eight} but for $\xi =1$. The thin solid (black) line in both panels labels the result of Ref.~\onlinecite{chen_prb_1999} [see Eq.~(\ref{eq:chenraikh})], which is asymptotically exact in the weak SOC limit.\label{fig:nine}}
\end{figure*}
\subsection{Taking into account the density dependence of the Rashba SOC}

Until now we have treated the Wigner-Seitz parameter $r_s$ (or density) and the Rashba SOC constant $\alpha$ 
as two {\it independent} parameters. This is similar in spirit to what has been done 
for decades in the context of tunnel-coupled double quantum wells where the single-particle symmetric-to-antisymmetric gap 
$\Delta_{\rm SAS}$ and density have been treated as independent parameters (see {\it e.g.} Ref.~\onlinecite{abedinpour_prl_2007} and references therein to earlier work). 

In reality, when a {\it single} gate voltage is applied to the 2DEG to change its density (and thus the $r_s$ value) the asymmetry of the quantum well which hosts the 2DEG changes too~\cite{twogates}. This in turn changes $\alpha$. In a simple single-band model with infinite barriers the SOC strength $\alpha$ is given by~\cite{knap_prb_1996,nitta_prl_1997,grundler_prl_1999}
\begin{equation}\label{eq:one-body-alpha}
\alpha = e \alpha_{\rm so} \langle E \rangle \approx \frac{e \alpha_{\rm so} n}{\epsilon}~,
\end{equation}
with $\alpha_{\rm so} = 117$~\AA$^2$ for bulk InAs. Here we have used that 
the electric field in the well is given by $ \langle E \rangle = n_{\rm d} /\epsilon$ where the  
density of the donors $n_{\rm d}$ has been approximated by the density of electrons $n$.

In Fig.~\ref{fig:ten} we present numerical results for the ratio ${\widetilde \alpha}_\pm/\alpha$ calculated by taking into account the density dependence of $\alpha$ according to Eq.~(\ref{eq:one-body-alpha}). From this plot we clearly see that the difference between the effective SOC constants ${\widetilde \alpha}_+$ and ${\widetilde \alpha}_-$ becomes negligibly small and that the ratio ${\widetilde \alpha}_\pm/\alpha$ changes by roughly sixty percent when density is changed over three orders of magnitude (for unscreened Coulomb interactions). The enhancement of SOC due to interactions increases with decreasing density. Overscreening Coulomb interactions washes out this effect yielding a tiny renormalization over the same density range.

\begin{figure*}
\centering
\includegraphics[width=0.48\linewidth]{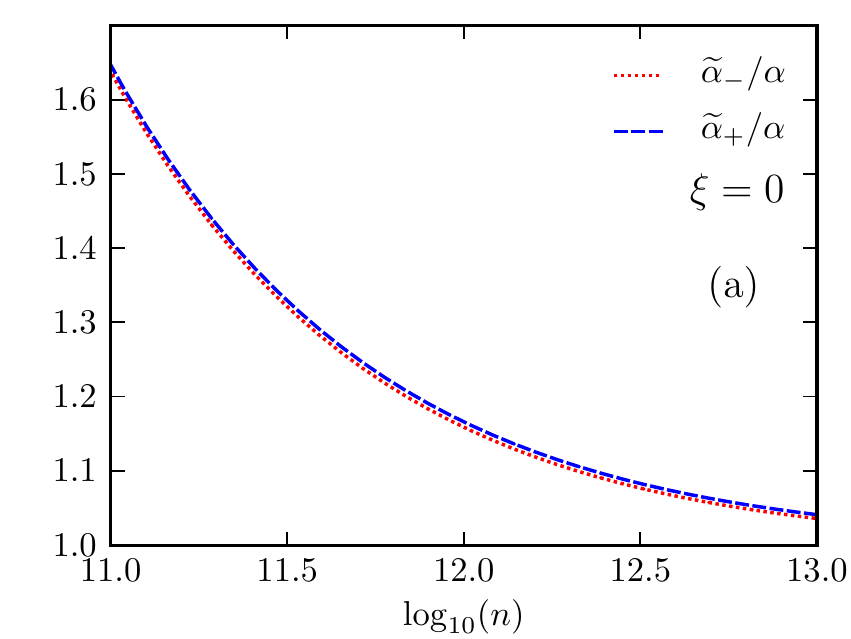}
\hspace{0.5cm}
\includegraphics[width=0.48\linewidth]{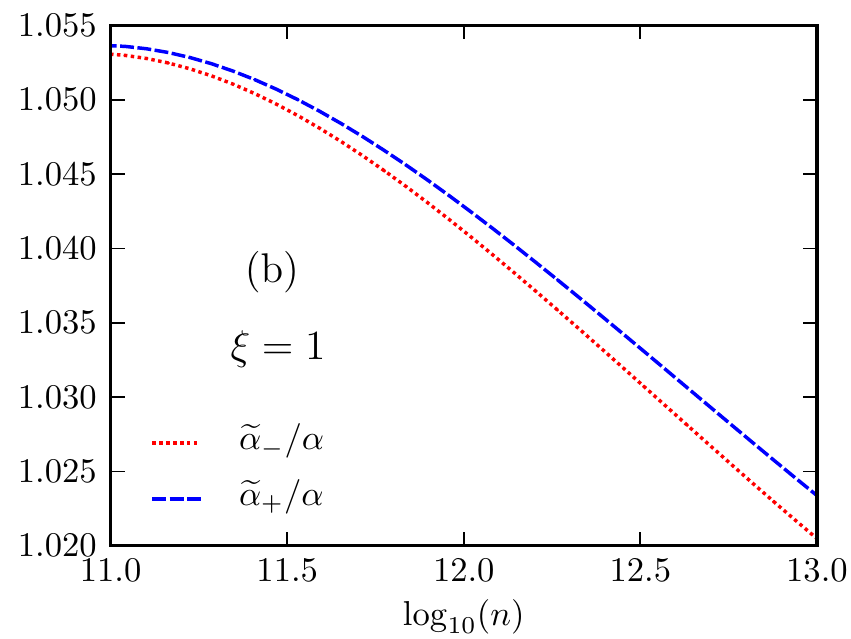} 
\caption{(Color online) The renormalized SOC constants ${\widetilde\alpha}_{\pm}$ [in units of the noninteracting value $\alpha$] are plotted as functions of the logarithm of density (expressed in units of ${\rm cm}^{-2}$). In this figure we have taken into account the density dependence of the bare SOC coupling $\alpha$ in the simple approximation given in Eq.~(\ref{eq:one-body-alpha}). Panel a) Results for unscreened Coulomb interactions ($\xi=0$). Panel b) Results for fully-screened Thomas-Fermi interactions ($\xi=1$). \label{fig:ten}}
\end{figure*}

Finally, in Fig.~\ref{fig:eleven} we show the spin susceptibility enhancement $\chi_{\sigma^y\sigma^y}/\chi^{(0)}_{\sigma^y\sigma^y}$ 
as a function of $r_s$ calculated by taking into account the dependence of the bare $\alpha$ on density {\it via} Eq.~(\ref{eq:one-body-alpha}). Note that for unscreened Coulomb interactions the enhancement can be as large as $20-30\%$ for $n\approx 10^{10}~{\rm cm}^{-2}$ (recall that in InAs $r_s =1$ corresponds to an electron density $\approx 2.6 \times 10^{10}~ {\rm cm}^{-2}$).

\begin{figure}
\begin{center}
\includegraphics[width=1.00\linewidth]{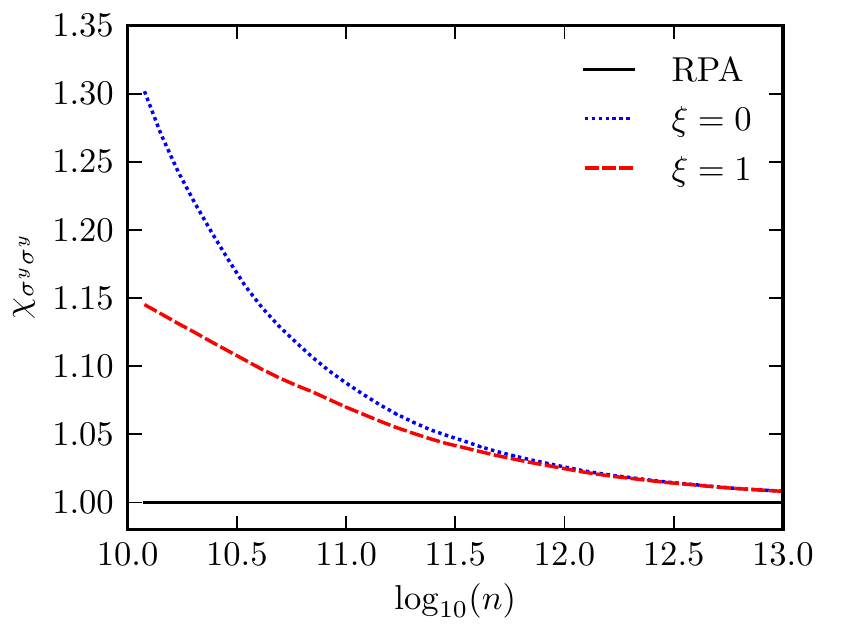}
\caption{(Color online) The spin susceptibility enhancement $\chi_{\sigma^y\sigma^y}/\chi^{(0)}_{\sigma^y\sigma^y}$ as a function of the logarithm of density (expressed in units of ${\rm cm}^{-2}$). In this figure we have taken into account the density dependence of the bare SOC coupling $\alpha$ in the simple approximation given in Eq.~(\ref{eq:one-body-alpha}).\label{fig:eleven}}
\end{center}
\end{figure}

\section{Conclusions}
\label{sect:conclusions}

In summary, we have studied the long-wavelength plasmon dispersion and the Drude weight of a two-dimensional electron gas with Rashba spin-orbit coupling. We have shown that these measurable quantities are sensitive to electron-electron interactions due to broken Galileian invariance and we have discussed in detail why the random phase approximation is not capable of describing the collective dynamics of these systems even at very long wavelengths. We have then presented approximate microscopic calculations of these quantities based on the so-called time-dependent Hartree-Fock approximation. We have found that interactions {\it enhance} the plasmon mass and suppress the Drude weight. 

These findings can in principle be tested experimentally by inelastic light scattering, electron energy loss, and far-infrared optical-absorption measurements. Inelastic light scattering~\cite{vittorio_ils} has already been extensively used to measure the plasmon dispersion in GaAs quantum wells~\cite{olego_prb_1982,hirjibehedin_prb_2002}. Notable deviations from the predictions of the random phase approximation have been observed~\cite{hirjibehedin_prb_2002} at finite momentum transfer $q$ and at low densities. We hope that similar studies can be performed systematically in asymmetric $n$-doped quantum wells with tunable spin-orbit coupling.

Last but not least, we have also computed quantitatively the renormalization of the Rashba spin-orbit coupling constant due to electron-electron interactions and the interaction corrections to the clean-limit spin Hall conductivity.

In the future we plan to extend these studies to the complete spin-orbit-coupling model Hamiltonian (\ref{eq:SOC}) with both $\alpha$ and $\beta$ finite. As already stressed in the main body of this article, this complicates things quite a bit since the resulting ground state does not possess rotational invariance. It is worth exploring also other spin-orbit-coupled two-dimensional quantum liquids such as 
hole gases in quantum wells which have a very rich single-particle band structure~\cite{winkler_book}. 

\acknowledgments 
A.A., T.J., and M.P. gratefully acknowledge funding from the European
Community's Seventh Framework Programme (FP7/2007-2013) under grant
agreement n. 215368 (SemiSpinNet). T.J. also acknowledges support 
from Czech Republic Grants AV0Z10100521, KAN400100652, LC510, and Preamium Academiae. 
J.S. was supported by NSF grant No. DMR-0547875 and by the Research Corporation Cottrell Scholar Award. 
G.V. was supported by the National Science Foundation under grant number DMR-0705460. 

\begin{thebibliography}{77}
\bibitem{footnote_crystal_field_effect}
	Strictly speaking, the low-energy Dirac-Weyl effective Hamiltonian describing electrons close to the Dirac point 
	in graphene~\cite{graphene_review} does not stem from relativistic effects but is purely an effect of the 2D periodic potential of the lattice. 
	The pseudorelativistic character of the low-energy Hamiltonian in graphene and the consequent lack of Galileian invariance 
	are merely a crystal field effect.
\bibitem{Hartree}
	 We are here dropping the so-called Hartree terms, which have ${\bm q}={\bm 0}$, 
	 and are thus canceled by the presence of the uniform background of neutralizing positive charge.
\end{thebibliography}
\end{document}